\documentclass[11pt,a4paper]{article}
\pdfoutput=1

\usepackage[utf8]{inputenc}
\usepackage[english]{babel}

\usepackage{extarrows}
\usepackage{amsmath}
\usepackage{amsfonts}
\usepackage{amssymb}
\usepackage{graphicx}
\usepackage{fourier}
\usepackage{authblk}
\usepackage{xcolor}

\usepackage{caption}
\usepackage{subcaption}
\usepackage{float}
\usepackage{appendix}
\usepackage{authblk}
\usepackage[left=2cm,right=2cm,top=2cm,bottom=2cm]{geometry}

\numberwithin{equation}{section}

\usepackage[hidelinks]{hyperref}

\newcommand\nn{\nonumber}
\newcommand\be{\begin{equation}}
\newcommand\ee{\end{equation}}
\newcommand\ba{\begin{eqnarray}}    
\newcommand\ea{\end{eqnarray}}      

\begin{document}

\title{Modular Hamiltonian for (holographic) excited states}

\author[a,b]{Ra\'ul Arias}
\author[b]{Marcelo Botta-Cantcheff}
\author[b]{Pedro J. Martinez}
\author[b]{Juan F. Zarate}
\affil[a]{SISSA and INFN, via Bonomea 265, 34136 Trieste, Italy}
\affil[b]{Instituto de F\'isica La Plata - CONICET and 

Departamento de F\'isica, Universidad Nacional de La Plata 

C.C. 67, 1900, La Plata, Argentina}

\maketitle

\begin{abstract}

In this work we study  the Tomita-Takesaki construction for a family of excited states that, in a strongly coupled CFT - at large $N$-, correspond to coherent states in an asymptotically AdS spacetime geometry. We compute the Modular flow and Modular Hamiltonian associated to these excited states in the Rindler wedge and for a ball shaped entangling surface. Using holography, one can compute the bulk modular flow and construct the Tomita-Takesaki theory for these cases. We also discuss generalizations of the entanglement regions in the bulk and how to evaluate the Modular Hamiltonian in a large N approximation. Finally, we extend the holographic BDHM formula to compute the modular evolution of operators in the corresponding CFT algebra, and propose this as a more general prescription.
\end{abstract}

\tableofcontents

\newpage

\section{Introduction}

Modular Hamiltonians (also called entanglement Hamiltonians in the condensed matter community) are unbounded and hermitian operators properly defined in the context of Axiomatic Quantum Field theories  \cite{Haag, Witten2018}, and in particular in the framework of the Tomita-Takesaki (TT) theorem \cite{Takesaki}.
However, they are also useful in other areas of Physics as Quantum Information, Quantum Field Theory and also in the AdS/CFT context.

Given a theory on a spacetime ${\cal{M}}$ in a state defined through its density matrix $\rho$ we can define the reduced density matrix, $\rho_{A}$, on a subsystem ${A}\in{\cal{M}}$ as the partial trace on the complement of $A$ (denoted by $\bar A$). By definition this object is semi-definite positive and Hermitian and then can be always written as

\begin{equation}
\rho_{A}=\text{Tr}_{\bar A}\rho=\frac{e^{-K_{A}}}{\text{Tr} e^{-K_{A}}},\label{defi}
\end{equation}
where $K_{A}$ is the Modular Hamiltonian. The denominator ensures $\text{Tr}\rho_{A}=1$ but will not play an important role along this work.

In the condensed matter literature, the spectrum of the Modular Hamiltonian is important because it have relevant information to characterize and identify topological states of matter. For example, in \cite{Li} it was applied to Fractional Quantum Hall states and, more recently, it was used to analyze topological non-hermitian systems \cite{Chang}. In
 QFTs, it is also an important tool when computing information theory measures, such as
the relative entropy 
between two states 
 \cite{RauloLocal} or the capacity of entanglement \cite{deboer} 
but since Modular Hamiltonians are typically non-local they are not easy to compute in general. Despite of this, its explicit form is known for the Rindler wedge in any QFT's vacuum state \cite{Bisognano}, as well as for a spherical entangling surface in CFTs \cite{Casini2011} and in time dependent situations after a quantum quench \cite{Cardy}.
 Some new analytical results were also found recently for free theories and multiple intervals on Minkoski spacetime \cite{Arias2018} and the torus \cite{BlancoGuillem}. Lastly, 
  we mention that Modular Hamiltonians have also been relevant in the context of the AdS/CFT correspondence
 \cite{adscft, GKP, W}, 
 in studying the Bekenstein bound \cite{Casinibound},
 the Averaged Null Energy (ANEC) and Quantum Null Energy (QNEC) conditions  \cite{Faulkner}, and emergent gravity \cite{An}. There is also a way to study Modular Hamiltonians in AdS/CFT on the gravity dual theory
 \cite{JLMS, Jafferis}.

The Tomita-Takesaki \cite{Takesaki} theorem is one of the most important theorems in the algebraic quantum field theory setup and despite being a very formal tool, it has many applications in physics and mathematics, see \cite{Witten2018} and references therein.
Let $\mathcal{A}$ be a von Neumann algebra on a certain Hilbert space $\mathcal{H}$ which contains a vector $|\Omega\rangle$ that is cyclic and separating on $\mathcal{A}$. Let us now define and operator $S$ on $\mathcal{H}$ by the following relation
\begin{equation}
S A |\Omega\rangle=A^\dagger|\Omega\rangle,\,\,\,\,\,\,\,\,\,\, \forall A\in {\cal{A}}.\label{Stt}
\end{equation}
The operator $S$ is called the Tomita operator with respect to $(\mathcal{A},\Omega)$ and it has a unique polar decomposition
\begin{equation}
S=J\Delta^{1/2}=\Delta^{-1/2}J,\label{tomita} 
\end{equation}
where the operator $\Delta=S^\dagger S$ is called the modular operator and the anti-unitary operator $J$ is called modular conjugation. From the definition can be seen that $J=J^\dagger$ and $J^2=1$. The TT theorem states that for $A$ a subalgebra of $\mathcal{A}$ and $|\Omega\rangle$ a separating and cyclic vector on $\mathcal{A}$ then the following properties holds

\begin{equation}
J|\Omega\rangle=\Delta|\Omega\rangle=|\Omega\rangle,\,\,\,\,\,\,\,\,\,\,\,\,\,\,\,
JAJ=A',\,\,\,\,\,\,\,\,\,\,\,\,\,\,\,
\Delta^{i s}A\Delta^{-i s}=A,\,\,\,\,\,\,\,\, \forall s\in \mathbb{R}.
\end{equation}
Here the subalgebra $A'$ belongs to the commutant of $\mathcal{A}$ (called $\mathcal{A}'$). This theorem ensures that there exist an uniparametric group of automorphisms $\sigma_s(A)=\Delta^{i s}A\Delta^{-i s}$ that will be called in our context as modular flow.  Usually it is non-local (non geometric) but there exist some important examples where it is local (geometric). Typical examples of local flows are those produced by modular operators in the vacuum state and for the Rindler wedge \cite{Bisognano}, or the vacuum of a CFT on a sphere \cite{Casini2011} and more recently the single interval case of fermions on a torus \cite{BlancoGuillem}. Typical examples of a non local flow is when the region is on the real line and made of disjoint intervals \cite{ Arias2018}  and as well when we study fermions in disjoint intervals on the torus \cite{BlancoGuillem}.

The Modular Hamiltonian $K_A$ is a self-adjoint operator that belongs to the algebra of operators, defined on the region $A$. It has a very precise definition in the AQFT setup in the context of the TT modular theory \cite{Takesaki}. However, it is often difficult to relate the modular operator with the one defined in equation \eqref{defi}. One of the goals of the present work is to compute $\Delta$ in a special class of excited states that have an holographic dual. Some recent works on the Tomita modular operator show its relevance in the derivation of
the ANEC \cite{Faulkner, Faulkner2} and 
in understanding aspects of black holes interiors \cite{Papadodimas2013, Jefferson2018} and bulk reconstruction \cite{Faulkner2,Czech2019, deBoer2019}. A remarkable feature of the Modular Hamiltonian is the fact that it depends only on the algebra of operators on the region of interest and the state of the system. Moreover, as was mentioned before, the TT theorem ensures that a notion of modular {\emph{time}} evolution can be defined. A local observer with a clock that runs in modular time will find himself in a thermal bath which is reminiscent of what happens in the Unruh effect.

In the present work we will study the Modular Hamiltonian, modular operator and its modular flow for a family of excited states in equipartite Hilbert spaces in the context of holographic CFTs.
Some related works in the field theory context are \cite{Lashkarimod, Sarosi, Pando} (see \cite{Liu} for an axiomatic approach). 
The particular set of excited states, that we will call \emph{holographic}, have the advantage that its precise holographic dual is known \cite{Peter 1, Skenderis} and  can be shown to be bulk coherent states on the large $N$ limit. 
Due to this important property, these states were extensively studied in different setups \cite{Marolf, Peter 2, Belin, Mark} and extended to finite temperature cases \cite{Peter 3, Peter 4}. To be concrete, the excited states are built by considering external sources in the Euclidean path integral that define a reference state, and can be written as (notation will be made explicit is Sec. \ref{states})
\begin{equation}\label{duality}
 |\Psi_\lambda\rangle \equiv {\cal P} \,e^{-\int_{\tau<0} d\tau \; {\cal O}(\tau) \cdot \,\lambda(\tau)}\,|0\rangle  \,\qquad\,\Leftrightarrow\,\qquad\, \langle\phi_{\Sigma}|\Psi_\lambda\rangle\equiv\int{\cal D}_{\phi_{\Sigma};\lambda}\Phi e^{-S_E[\Phi]}
\end{equation} 
where the object on the left is an excited state on the CFT, and the object on the right is its wave function in the holographic dual.

In section \ref{states} we will review the main properties of the states \eqref{duality} that will be useful
in the context of our work.
In section \ref{modularqft} we will study the Modular Hamiltonian of these \emph{excited}-states from a QFT perspective for the case of an equipartite system. 
Throughout this section, we will emphasize on the TT theory and the geometric structure associated to it that follows from our analysis. For a CFT, this analysis can be extended to
other partitions via conformal maps.
The main explicit computations are presented in Sec. \ref{gravity} for a free scalar field in the gravity dual, where by virtue of the large-N approximation, the problem with general $\tau$-dependent $\lambda$ becomes tractable and,
 by using Thermo-Field Dynamics (TFD) tools, we find the explicit details of the TT construction in the gravity dual. The underlying proposal is that holographic methods can be used to treat some formal as well as quantitative aspects related with the modular theory. 
In particular, using the BDHM recipe, we compute the excited modular flow for scalar operators in the field theory from the explicit calculation in the bulk.
In section \ref{conclu} we summarize our results and discuss some open problems.
We also include two Appendixes: App. \ref{tfd} contains a brief introduction to the relevant aspects and tools of the TFD formalism and in App. \ref{App:Ball} we explicitly show the extension of our results to a spherical entangling region using the Casini-Huerta-Myers map \cite{Casini2011}.

\section{Excited states and holographic dictionary}\label{states}

In this section we briefly summarise some relevant properties and known results on excited states in the holographic context that will be the basis of the present work. A prescription for holographic states appear in the Skenderis-van Rees works \cite{SvRC,SvRL} which combines Lorentzian and Euclidean AdS spacetimes, as being dual to real and imaginary time pieces of a Schwinger-Keldysh path in the field theory (e.g. see Fig \ref{Fig:Camino}). Such as in the Hartle-Hawking (HH) formalism \cite{HH}, the imaginary time intervals describe \emph{states}, which are connected by evolution on the real-time intervals. 
The excited states 
are deformations of the vacuum HH state related to non vanishing asymptotic (Euclidean) boundary conditions, and have been recently studied in many different contexts \cite{Peter 1,Peter 2, Mark,Peter 4, ensayo-us}.

\subsection{Holographic states: definition and geometric motivation}

Consider a CFT defined by a conformally invariant action, on a spacetime $(\mathbb{R}^{d+1} , \eta_{\mu \nu})$. In the interaction picture the states defined in \cite{Peter 1} have the form
\begin{equation}\label{ket-IP}
 |\Psi_\lambda\rangle \equiv {\cal P} \,e^{-\int_{\tau<0} d\tau \; {\cal O}(\tau,x^i) \cdot \,\lambda(\tau,x^i)}\,|\Psi_0\rangle.
\end{equation}
Here, $|\Psi_0\rangle$ denotes the vacuum state and $\tau$ represents the Euclidean time, and $\cdot$ denotes integration on the spatial coordinates on a certain Cauchy surface. The function $\lambda(\tau, x^i)$ parameterizes the family of states, and can be arbitrarily chosen on the asymptotic boundary of (a half of) an Euclidean aAdS spacetime (called $E^- $). The object ${\cal O}$ is an operator{\footnote{Originally in \cite{Peter 1} only single trace operators were considered. Recently, \cite{Vanraam2019} studied the effect of multi-trace deformations.} of the CFT and the operation denoted by ${\cal P}$ denotes
the path ordering in the (imaginary) time. 
We will often omit the $x^i$ dependence of ${\cal O}$ and $\lambda$. By extending the source $\lambda^\star(\tau) \equiv \lambda(-\tau)$ to the region $0<\tau<\infty$, one obtains the corresponding ''bra'',
 \begin{equation}\label{bra-IP}
\langle\Psi_\lambda| \equiv \langle \Psi_0| \, {\cal P} \,e^{-\int_{\tau>0} d\tau \; {\cal O}(\tau,x^i) \cdot \,\lambda^\star(\tau,x^i)}\,.
\end{equation}
According to the Hartle-Hawking construction, if $\lambda$ is independent of $\tau$, the states \eqref{ket-IP} can also be thought as the ground state of a deformed Hamiltonian \cite{Belin},
\be
H=H_0 + 
\int dx \;
{\cal O} (x,0)\,\lambda(x)\;.
\ee
On the other hand, if the source  $\lambda=\lambda(\tau, x)$ is vanishing when $\tau \to 0$ and decays to zero as $\tau \to -\infty$, the state \eqref{ket-IP} can be interpreted as the result of a (Wick rotated) time evolution of the fundamental state perturbed by an external source, i.e. in Schrodinger picture,
\begin{equation}\label{state-open}
 |\Psi_{\lambda}\rangle =  U_{\lambda} |\Psi_0\rangle= {\cal P} \,e^{-\int_{\tau<0} d\tau \; (H + {\cal O}. \,\lambda(\tau))}|\Psi_0\rangle\;.
\end{equation}
This procedure provides excited states of the original theory \cite{Glauber}.
These states are specially interesting in the context of holography, where there is a precise (non perturbative) prescription for the states \eqref{ket-IP} and its corresponding bra \eqref{bra-IP} in the dual bulk theory:
 \be\label{state-pathI}
 \Psi_\lambda [\phi_\Sigma] = \langle\phi_\Sigma|\Psi_\lambda\rangle = \int_{\Phi|_{\partial {\cal E}^-}=\lambda\,,\,\Phi|_{\Sigma}=\phi_\Sigma} [D\Phi] \; e^{-S[\Phi]}\;.
 \ee
 Here ${\cal E}^-$ is an Euclidean spacetime whose metric is locally AdS on the asymptotic boundary $\partial {\cal{E}}^-=E^-$, $\Sigma$ is the spacelike surface over which the state is defined,
  and $S[\Phi]$ stands for the bulk gravity action, say for a bulk scalar field $\Phi$, dual to the scalar (primary) operator ${\cal O}$ in the CFT theory.
   Implicitly this action contains an Einstein-Hilbert term proportional to $G_N^{-1} \sim N^2$ that will contribute to the Ryu-Takayanagi area term, see Sec. \ref{Sec:AdS-CFT}.
     In the present study we will focus on the subleading ($\propto G_N^0$) terms, coming from the matter sector and backreaction effects, that shall describe the quantum corrections to the (bulk) Modular Hamiltonian and entanglement entropies \cite{Aitor2013}.  The expression \eqref{state-pathI} can be derived from the Skenderis - van Rees (SvR from now on) proposal \cite{Peter 1,SvRC,SvRL}, and generalizes the Hartle-Hawking wave functional of the gravitational vacuum to excited states. In what follows we will refer to these CFT states \eqref{ket-IP} simply as \emph{holographic states}. We want to stress that there is a simple holographic dictionary that characterize them: \emph{Deformations of the  CFT action on the euclidean times $\tau<0$ correspond to deformations of the (Dirichlet) boundary conditions for the bulk fields on the Euclidean section of the dual aAdS spacetime.} 
Notice that this rule captures the holographic correspondence between the Hartle-Hawking vacua of both (CFT and gravitational) theories in absence of sources ($\lambda\to 0$). As a by-product, in the large $N$ limit, as gravity becomes semiclassical, the vacuum is given by a unique (classical) Euclidean aAdS spacetime which corresponds to the CFT vacuum state $|\Psi_0\rangle$. 
This prescription will allow to obtain the holographic dual of the modular flow for arbitrary holographic states and arbitrary regions $A$ of a Cauchy slice $t=0$ of the boundary spacetime $\partial {\cal E} $.

\begin{figure}
  \centering
\begin{subfigure}{0.49\textwidth}\centering
\includegraphics[width=.9\linewidth] {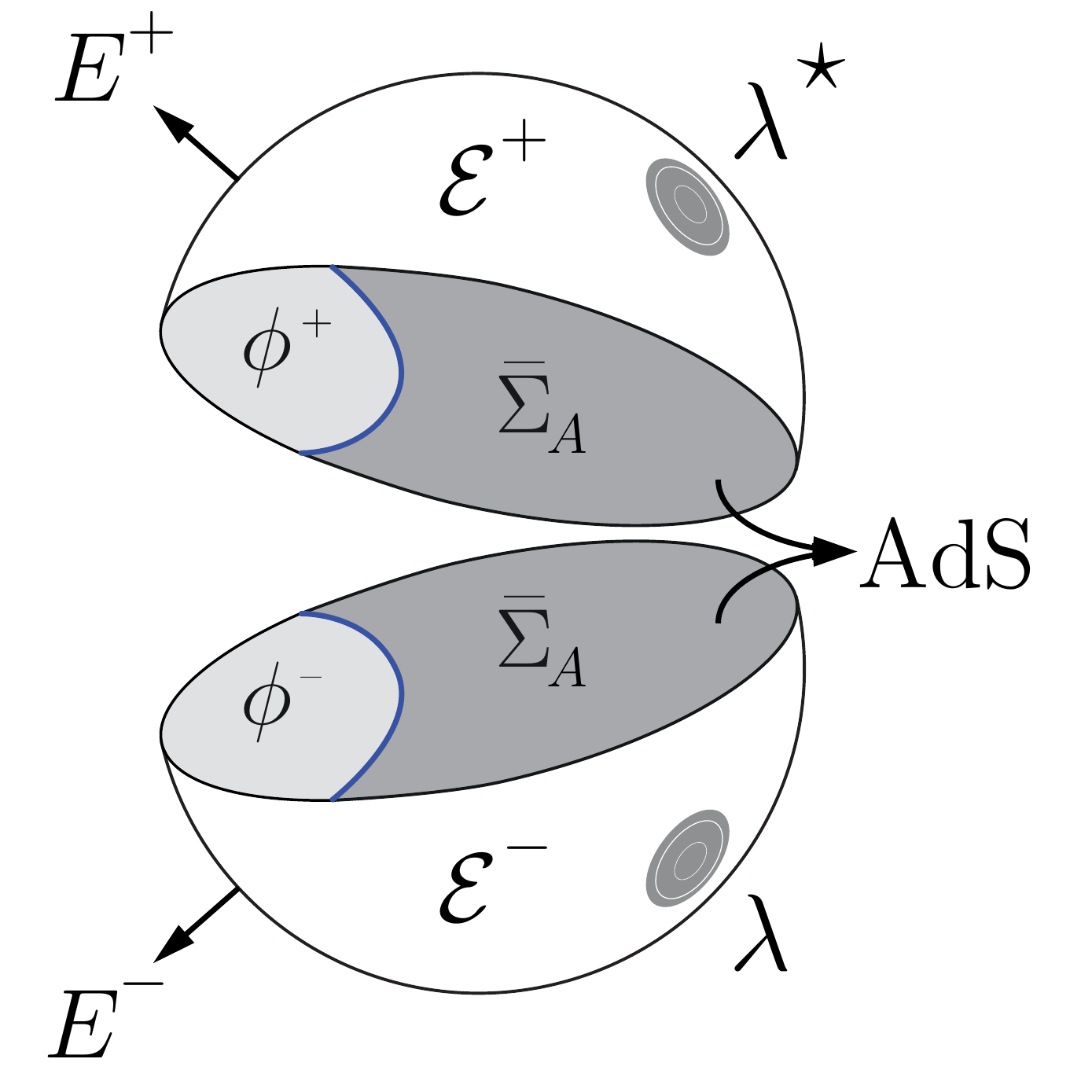}
\caption{}
\end{subfigure}
\begin{subfigure}{0.49\textwidth}\centering
\includegraphics[width=.9\linewidth] {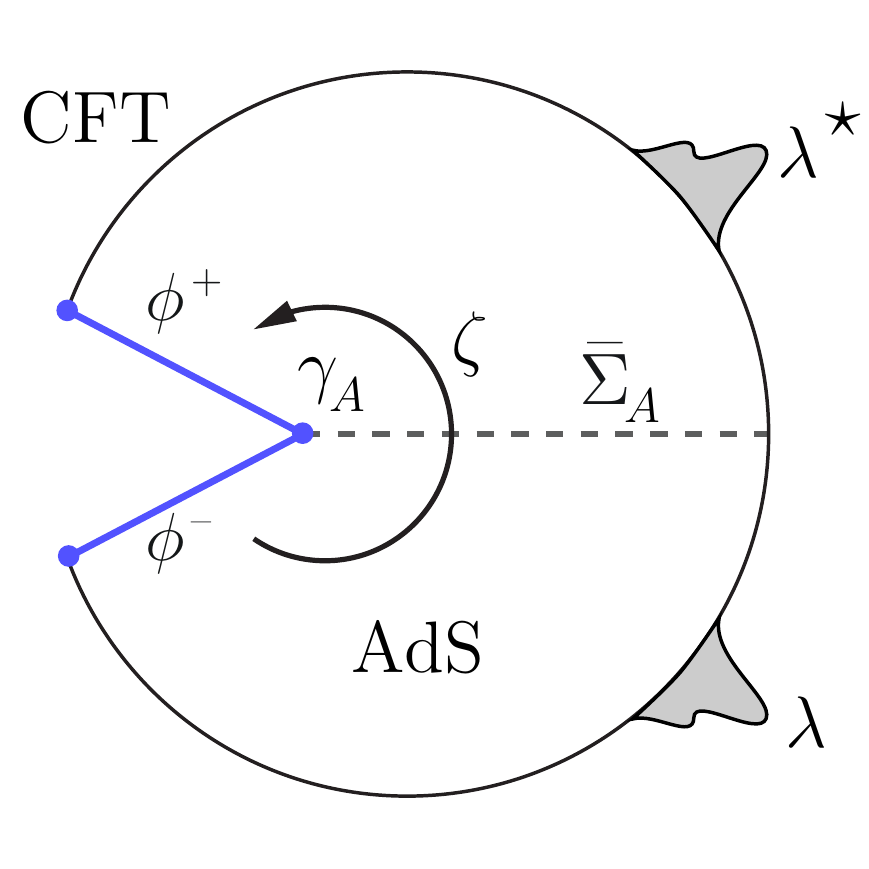}
\caption{}
\end{subfigure}
  \caption{A depiction of the computation of $Tr_{\overline{\Sigma}_A}\,|\Psi_\lambda\rangle\langle\Psi_\lambda |$  in \eqref{density-bulk} is shown. The figure in the left represents the gluing of the states, \eqref{ket-IP} and \eqref{bra-IP}, by tracing over $\bar \Sigma_A$, the complement of $\Sigma_A\subset\Sigma$, shown in dark grey. The CFT external source $\lambda$, as well as conditions $\phi^{\pm}$, provide boundary conditions for the path integral in the bulk. The blue line corresponds to the extremal surface $\gamma_A$. On the right, an example where a $\zeta$ Killing vector that runs as an angle from $\Sigma_A^-$ to $\Sigma_A^+$ pivoting around the blue point $\gamma_A$ is shown. All but radial and euclidean time coordinates have been suppressed.}
  \label{fig:evo2}
\end{figure}

Another important feature is that with these kind of states we can study density matrices and reduced density matrices in a clean way. Considering \eqref{ket-IP}, we can define the density matrix reduced to the subsystem $A$ as
\be\label{density-bulk}
 \rho_\lambda [\phi^+, \phi^-]\equiv  \langle \phi^+ |\rho_\lambda | \phi^-\rangle=\int_{\lambda, \phi^\pm} {\cal D}\Phi \; e^{-S[\Phi]}\;,
 \ee
 where $\phi^\pm\equiv \phi|_{\Sigma_A}$, represents two different data on the surface $\Sigma_A$.
  Additionally, we have to impose the asymptotic conditions
$\Phi|_{\partial {\cal E}^-}=\lambda\,, \Phi|_{\partial {\cal E}^+}=\lambda^\star $, and vanishing data for all the other bulk fields (including the graviton).
 This expression is straightforwardly obtained by gluing two halves of Euclidean AdS on the shaded regions of figure \ref{fig:evo2}, which represents the operation $ Tr_{\overline{\Sigma}_A}\,|\Psi_\lambda\rangle\langle\Psi_\lambda | $ in a path integral description.

The Ryu-Takayanagi prescription (RT) determines the entangling surface in the bulk $\gamma_A$, so as the spatial region $\Sigma_A$ delimited by $\gamma_A\, \cup\, A$. In fact, this formula should be interpreted in the sense of a perturbative expansion in the Newton constant $G_N$, 
so that the surface $\gamma_A$ (and $\Sigma_A$) remains unchanged by the back-reaction effects to each order.
In Sec. \ref{gravity} we will see that this prescription allows to obtain, at least formally, the holographic dual of the Modular Hamiltonian (and its corresponding modular flow) for holographic states and arbitrary regions $A$ of a Cauchy slice $t=0$ of the boundary spacetime $\partial {\cal E}$.

\subsection{Relation to coherent states in the bulk}

It has been stressed that states of this form are holographic in the sense that correspond to well defined geometric duals \cite{Marolf,ensayo-us,Faulkner2017}.  
One of the most interesting features of \eqref{state-open} is that, by canonically quantizing a (nearly) free non-backreacting field $\Phi$ in the bulk, these states become coherent in the large $N$ Hilbert space \cite{Peter 1}
\begin{equation}\label{coherent-state}
|\Psi_{\lambda}\rangle \propto e^{\int dk\; \lambda_k a^\dagger_k}|\Psi_0\rangle\,,
\end{equation}
where $a_k\,( a^\dagger_k\,)$ are the annihilation (creation) operators associated to the canonically quantized bulk field $\hat{\Phi}$ and  $\lambda^{}_k$ are eigenvalues of $a_k$, 
given by the Laplace transform of the Euclidean sources. 
This result can be achieved by using the so-called Banks, Douglas, Horowitz and Matinec (BDHM) prescription that relates the CFT with bulk field operators \cite{BDHM}, i.e, 
the operators ${\cal O}$ \eqref{state-open} are linearly expanded in terms of $a_k\, , \, a^\dagger_k$.

One can generalize this formalism to other more complex SK paths as we will see below, but the case presented here captures the essential aspects of this family of excited states. For example, by working in the Thermo Field Dynamics (TFD) formalism one can extend the definition of these states also to closed SK paths. This was done in \cite{Peter 3, Peter 4} and in what follows we will use some of their results.

An immediate application of \eqref{coherent-state} is that, in the free field approximation, these states work as \emph{generating} states since by simple derivation with respect to the normal modes components $\lambda_{w}$ we can obtain the expansion in a Fock basis. Although coherent states are an over-complete basis, they are associated to (generate) a complete orthogonal basis of the Fock space. Schematically,

\be\label{Fock-lambda}
|\Psi_\lambda\rangle \equiv \,  \prod_w \, \,e^{-\frac{|\lambda_w|^2}{2}}\, \sum_n  \,\frac{(\lambda_{w})^n}{\sqrt{n!}} (a_w^\dagger)^n |\Psi_0\rangle\;.
\ee
The product is on the (positive) frequencies $w$ of the normal-modes, and $\lambda_{w}$ are the components in the basis of functions on the Euclidean boundary induced from the normal-modes in the asymptotic boundary. 
The remarkable aspect of this expansion is that formally expresses the holographic state as a linear combination of operators on the ground state, that according to the BDHM dictionary \cite{BDHM}, can be translated to the CFT basis of states, $:({\cal O}_w)^n : |0\rangle$, where ${\cal O}_w$ are nothing but the (normal) frequency components of the local primaries ${\cal O}(\tau)$ on the Euclidean time $\tau<0$ \cite{Peter 1,Peter 2,Peter 4}. 

\section{Modular Hamiltonian for excited states in CFT}\label{modularqft}

In this section we will study the Modular Hamiltonian in the excited  states introduced in the previous section from a field theory point of view. In order to do so we will consider that these states lives in a bigger TFD Hilbert space. In Appendix \ref{tfd} we review TFD definitions and notations that will be useful.
We will start studying the situation where we can take the subsystem and its complement in an equipartite way i.e. the same number of operators on each of them. An example of this is the Rindler wedge, where we can think the total Hilbert space as ${\cal H}_{Tot}={\cal H}_{L}\otimes{\cal H}_{R}$ with ${\cal A}_{L,R}$ denoting the algebra of operators on the left (right) wedges respectively. But, the situation considered here will be more general than equipartite systems and then the results can be extended to any subregion that can be obtained from a conformal map from the Rindler result. As an example of this in App. \ref{App:Ball} we will follow the CHM map \cite{Casini2011} to obtain the Modular Hamiltonian in holographic excited states for a spherical entangling surface.

\subsection{Results in QFTs for equipartite subsystems } \label{Sec:sSK}

 Let us apply the construction reviewed in Sec. \ref{states} for a QFT defined on a globally hyperbolic spacetime $M\sim \Sigma \times \mathbb{R}$ with a Lorentzian metric ${\bf \eta}$ that we assume flat for simplicity. Consider first the case of an equipartition of the degrees of freedom in two equal sides $\Sigma \equiv \Sigma_L \cup \Sigma_R$, and the causal domain of both sides are called $W_{L/R}$ respectively. 
The equipartition requirement is not essential to study entanglement, but here we are motivated by an ingredient of the TFD formalism which supposes that the system described in $W_L$ (and the corresponding algebra of operators) is a \emph{copy} of the system that lives on $W_R$. 
In the Rindler example, the gravity dual has also two wedges connected by the horizon of a black hole, but there are also examples where each side could be compact and disconnected to each other, e.g. $\Sigma_{L/R} \sim S^{d-1}$, which is holographically related to an extended black hole  geometry \cite{Peter 3,Peter 4,eternal}. The results of this section apply to both possibilities. 
 If we consider the algebra of operators ${\cal A}$ restricted only to one wedge, say $W_R$, all the expectation values in a pure state $|\Psi_\lambda\rangle$ can be computed through a reduced density matrix $\rho_\lambda(R)$ by
 \be\label{n-points}
 Tr_R\, \{\rho_\lambda(R)\, {\cal O}(X_1){\cal O}(X_2) \dots {\cal O}(X_n)\} =  \langle \Psi_\lambda| \, {\cal O}(X_1){\cal O}(X_2) \dots {\cal O}(X_n) \,| \Psi_\lambda\rangle, \;\;\;\; \forall X_i \in \Sigma_R.
\ee
The vacuum state ($\lambda=0$) is thermal with respect to 
the Hamiltonian $K_0(R) \equiv -\log \rho_0(R)$, that coincides with the generator of the time translations for accelerated observers \cite{Bisognano}.
 
In the context of the present work, it will be useful to consider the \emph{symmetric} Schwinger-Keldysh (sSK) formalism and the extension of the Rindler time parameter to a \emph{closed} time contour ${\cal C}$ in the complex plane \cite{Peter 3,Peter 4}. The left wedge of the Minkowski spacetime $W_L$ can be identified with the backwards real-time component of the SK contour and the corresponding algebra of operators with $\widetilde{{\cal A}}$, the commutant of ${\cal A}$. The initial (and final) pure global state is described by the Euclidean intervals, see figure \ref{Fig:Camino}. The symmetric SK path shown in figure \ref{Fig:Camino}(a), that involves two imaginary path of equal length $\beta/2$ is \emph{equivalent} to the TFD formalism \cite{Schwinger,Keldysh,UmezawaTFD,UmezawaTFD2,UmezawaInteraction}.

\begin{figure}[t]\centering
\begin{subfigure}{0.49\textwidth}\centering
\includegraphics[width=.9\linewidth] {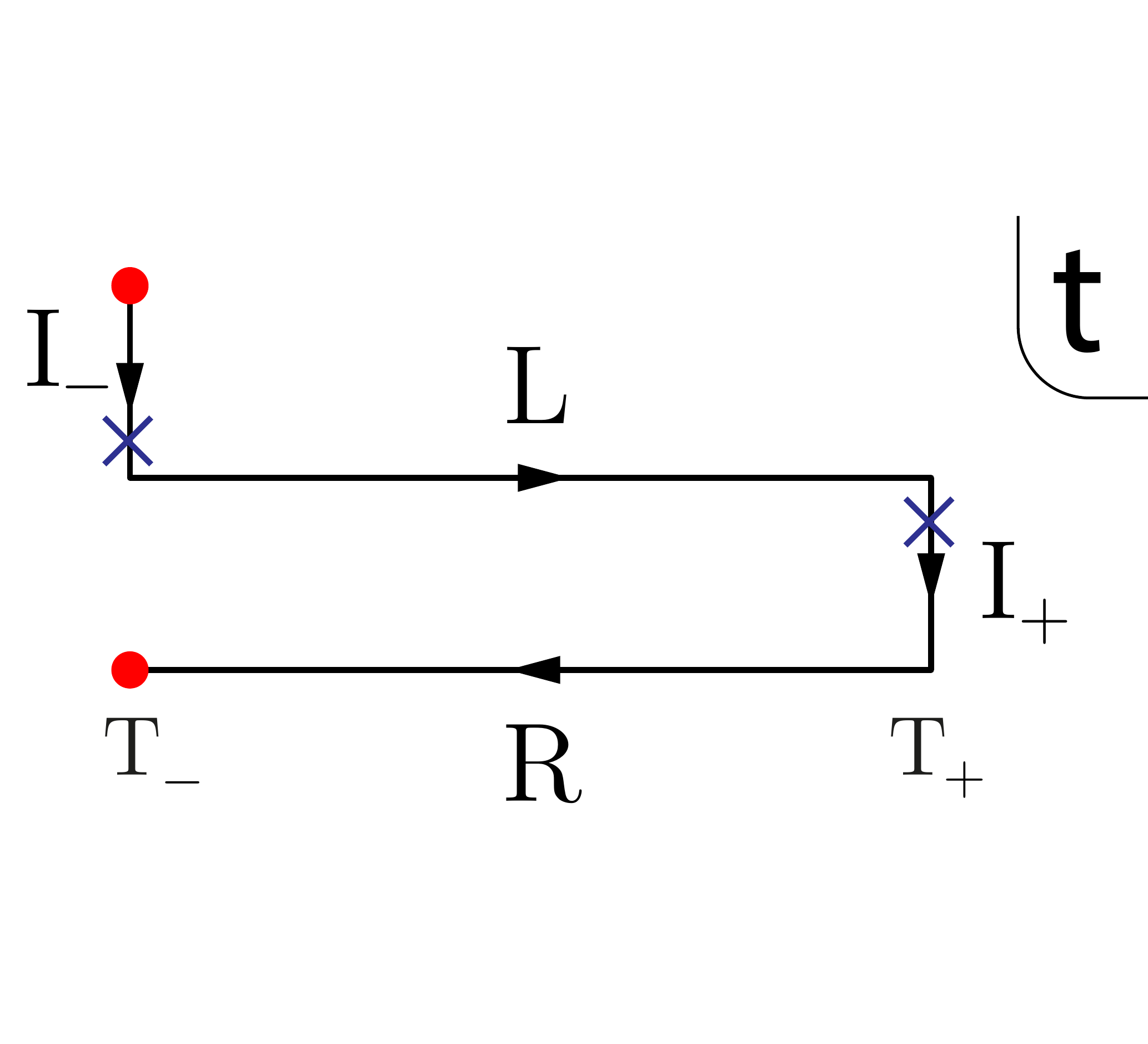}
\caption{}
\end{subfigure}
\begin{subfigure}{0.49\textwidth}\centering
\includegraphics[width=.9\linewidth] {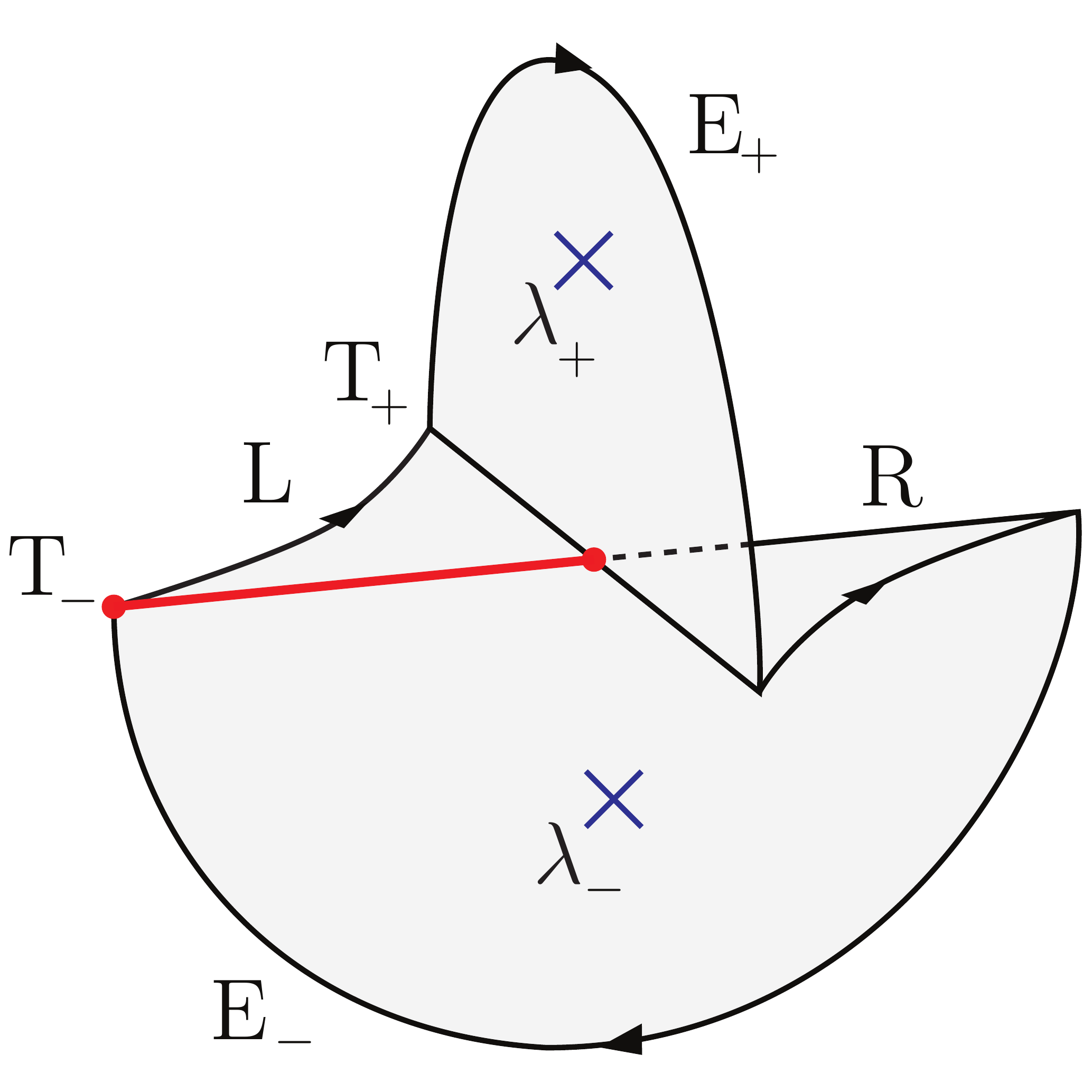}
\caption{}
\end{subfigure}
\caption{(a) Closed symmetric Schwinger-Keldysh (sSK) path in the complex $t$-plane.  The horizontal lines represent real time evolution. The vertical lines give imaginary time evolution, and the intervals $I_\pm$ have identical lengths equal to $\beta/2$.  The insertion of sources in the vertical lines generate excitations over the (vacuum) thermal state. (b) The radial Rindler coordinate is shown so that one can see that the L and R pieces are connected. Sources $\lambda_\pm$ are turned on in the corresponding Euclidean regions $E_\pm$. Notice that the red point in (a) is now a red line, representing a spacelike $\Sigma$ surface. The black arrow represents the path ordering of the operators given by ${\cal P}$, according to the parameter of the curve ($\theta$); it should not be confused with the time direction; for instance on the right side, the time parameter grows against the sense of the arrow (recall that $T_- < T_+$).} 
\label{Fig:Camino}
\end{figure}

 Denoting $X^\mu \in  W_R$ any point of the right wedge on the Rindler spacetime and using the notation $X^0 \equiv t$ and $X^d \equiv r$, the flat metric in conventional Rindler coordinates writes
\be\label{RindlerCFT} 
ds^2 = - \frac{r^2}{R^2} dt^2 + dr^2 + dX^i dX_i\,,\;\; i=1\dots d-1\,
\ee
and then, the modular flow for the vacuum state is
\be\label{evols}
 e^{-is K_0}{\cal O}_R(X^\mu) e^{is K_0} \equiv {\cal O}_R( \gamma^\mu (s)) = {\cal O}_R(r, X^i , t + s)\,,
\ee
with $\gamma(t)$ denoting the curve referred as geometric flow and ${\cal O}_R \in W_R$. Therefore a constant $t$ defines a particular foliation of $W_R$ in surfaces $\Sigma_R(t)$ that are homologous to $\Sigma_R$.
 We will compute the Modular Hamiltonian $K_\lambda$ for a thermal (in the sense mentioned before) excited state using the tools and ingredients of the TFD formalism, explained in App. \ref{tfd}. 
 
The evolution operator in this case is generated by  the CFT Hamiltonian $H\equiv{K_0}$ slightly deformed by external (local) sources $\lambda(x,\tau)$.
The operators $U_{\pm}$ on both imaginary time intervals, univocally describe the initial/final excited states in terms of 
$\lambda$ \cite{Peter 4}:
\begin{equation}\label{Upm}
 U_{\lambda} \equiv {\cal P} \,e^{-\int_{I_-} d\tau \; (K_0 + {\cal O}. \,\lambda(\tau))}\,. 
\end{equation}
where the Euclidean time $\tau$ runs on the interval $I_-\equiv(-\beta/2, 0)$, that alongside  $I_+\equiv(0,\beta/2)$ completes a (not closed) circle $S^1_\beta$ of radius $\beta$.
In the context of TFD these operators are equivalent to pure states, rearranged as \textit{kets} in the duplicated space.

Therefore, the global state is given in terms of this operator by \cite{Peter 3} (see Appendix \ref{tfd})
\be\label{state-U}
|\Psi_\lambda \rangle \equiv (U_\lambda \otimes \mathbb{ I}) |1\rangle\!\rangle  = (\mathbb{ I} \otimes U_\lambda) |1\rangle\!\rangle = U_\lambda \,|1\rangle\!\rangle ,
\ee 
and one can define an Hermitian (reduced) density matrix as
\be
\label{densitystate}
\rho_\lambda \equiv \text{Tr}_{\widetilde{{\cal A}}}\;\,|\Psi_\lambda \rangle \langle \Psi_\lambda |= Tr_{\widetilde{{\cal A}}}\,\,\,U_\lambda \, |1\rangle\!\rangle \langle\!\langle  1|\, U_\lambda^\dagger = \,U_\lambda\, \,U_\lambda^\dagger\,,
\ee
where we have used
\be
\text{Tr}_{\widetilde{{\cal A}}}\,\,\, |1\rangle\!\rangle \langle\!\langle  1|\,= \sum_n \,\, |n\rangle \langle n| = \mathbb{I}_{{\cal A}}\,.
\ee
These expressions explicitly show the connection between the pure state \eqref{state-U}  in the TFD setup and the mixed density matrix \eqref{densitystate}   in a single Hilbert space, both univocally determined by the evolution operator $U_\lambda$. Notice also that we can move the operator $U_\lambda$ back and forth between $L$ and $R$ when applied to $|1\rangle\!\rangle$. This will be a key property of the states \eqref{state-U} in our work. The main result of this section is that 
 for a thermal excited state \eqref{state-U} one can find that the reduced density matrix and Modular Hamiltonian on $W_R$ are
\be\label{density-rindler}
\rho_\lambda = {\cal P} \,e^{-\int_{S^1} d\tau \; (K_0 + {\cal O}. \,\lambda(\tau))}\,\qquad\qquad K_\lambda=-\ln (\rho_\lambda)=-\ln \{{\cal P} \,e^{-\int_{S^1} d\tau \; (K_0 + {\cal O}. \,\lambda(\tau))}\}
\ee
Although this is non-local by virtue of the integration on the interval $S^1$, it has a very simple form. 
Using eq. \eqref{densitystate} this result can be rephrased as 
\be\label{density-rindler-U}
\rho_\lambda = U_\lambda (-\pi , \pi) \, 
\ee
where the r.h.s. is the evolution operator valued on a imaginary time interval that  covers the complete circle $S^1$, provided that the source satisfies $\lambda(\tau) = \lambda(-\tau)$.

Recall that this
 captures all the excited states of the Hilbert space (often referred to as the vacuum  sector of the Hilbert space \cite{Witten2018}) defined as 
${\cal H}_0 \equiv {\cal A}| 0\rangle $, see \eqref{Fock-lambda}. The holographic dual of this space is the Fock space associated to quantized fields
in the bulk spacetime \cite{BDHM,BDHM2}. So, for instance, the unnormalized reduced density matrix corresponding to a single-particle state created at the point $(-i\tau , X) \in E^-$ ($0\leq \tau \leq \pi$), is 
\begin{align}\label{one-p-density-rindler} 
\rho_1 &= 
U_0 (-i\pi , - i\tau ) {\cal{O}}(-i\tau , X)  U_0  (-i\tau , 0 )\;
 U_0  (0, i\tau){\cal{O}}(i\tau , X) U_0  ( i\tau, i\pi)\nn  \\
 &= e^{(\tau -\pi) K_0} {\cal{O}}(-i\tau , X) e^{-2\tau K_0} {\cal{O}}(i\tau , X) e^{(\tau -\pi) K_0}
\end{align}
where $X^\mu \equiv (-i\tau, X)$ denotes a point on the surface $\Sigma_R (-i\tau)$ of the foliation of $E^-$. This expression can be derived from \eqref{Fock-lambda}, \eqref{state-U} and \eqref{densitystate} since in the sSK extension of the Rindler spacetime, the pure state writes\footnote{For a $n$-particle state we have to take $n$ derivatives.} 
\be \label{one-p-ket-rindler} 
|\Psi_1 \rangle = \left.\frac{\delta |\Psi_\lambda \rangle}{\delta \lambda(\tau , X)} \right|_{\lambda=0} = U_0 (-i\pi , - i\tau ) {\cal{O}}(-i\tau , X)  U_0  (-i\tau, 0 ) |1\rangle\!\rangle 
\ee
and using that ${\cal{O}}^\dagger(-i\tau , X) = {\cal{O}}(i\tau , X)$ and $U^\dagger(i\tau_i , i\tau_f ) = U(-i \tau_f , -i \tau_i)$.
In a CFT, expression \eqref{one-p-density-rindler} for the unnormalized density matrix can be generalized 
to \emph{any other region} conformally related to the Rindler wedge, say $D$: a ball shapped region (see App. B),
by inserting the (conformal) prefactor $\Omega^{-\Delta} (x(-i\tau)) \Omega^{-\Delta} (x(i\tau))$. Here $x(i\tau)$ denotes the geometric flow in the transformed space $D$, such that $x(0)$ stands for the conformal map of the point $(0 , X) \in \Sigma_R(0)$. Systematically, one could follow the same procedure for the n$^{th}$-order in a Taylor's  expansion in $\lambda$.

The main expressions \eqref{density-rindler}-\eqref{density-rindler-U} are non-perturbative result in the sense that holds for any $\lambda$. Upon completion of this work, we became aware of \cite{Parrikar2020}, where the authors did a perturbative analysis (similar to the first order \eqref{one-p-density-rindler}) of this result to compute the Modular Hamitonian in the states $|\Psi_\lambda \rangle$. 

A special case to be considered is whether the source $\lambda$ does not depend on $\tau$.
Thus, since the path ordering play no role, the (local) Modular Hamiltonian on $W_R$ results
\be\label{KR} 
K_\lambda \equiv 2\pi ( K_0 + {\cal O} \cdot \,\lambda)=    K_0 + \,\int_{\Sigma_R} \lambda( X) {\cal O}(X) \;\sqrt{g_{\Sigma_R}}dX^{d-1},\,
\ee
where $K_0$ can be expressed in terms of the energy-momentum tensor, the (timelike) Killing vector $\tau^\mu$ and the unit $n^\mu$ future pointing normal to $\Sigma_R$,
\begin{equation}
K_0=\int_{\Sigma_R} \, T^{\mu \nu }n_\mu \tau_\nu \;\,\sqrt{g_{\Sigma_R}}\,dX^{d-1}\;.
\end{equation}

\begin{figure}[t]\centering
\includegraphics[width=.65\linewidth] {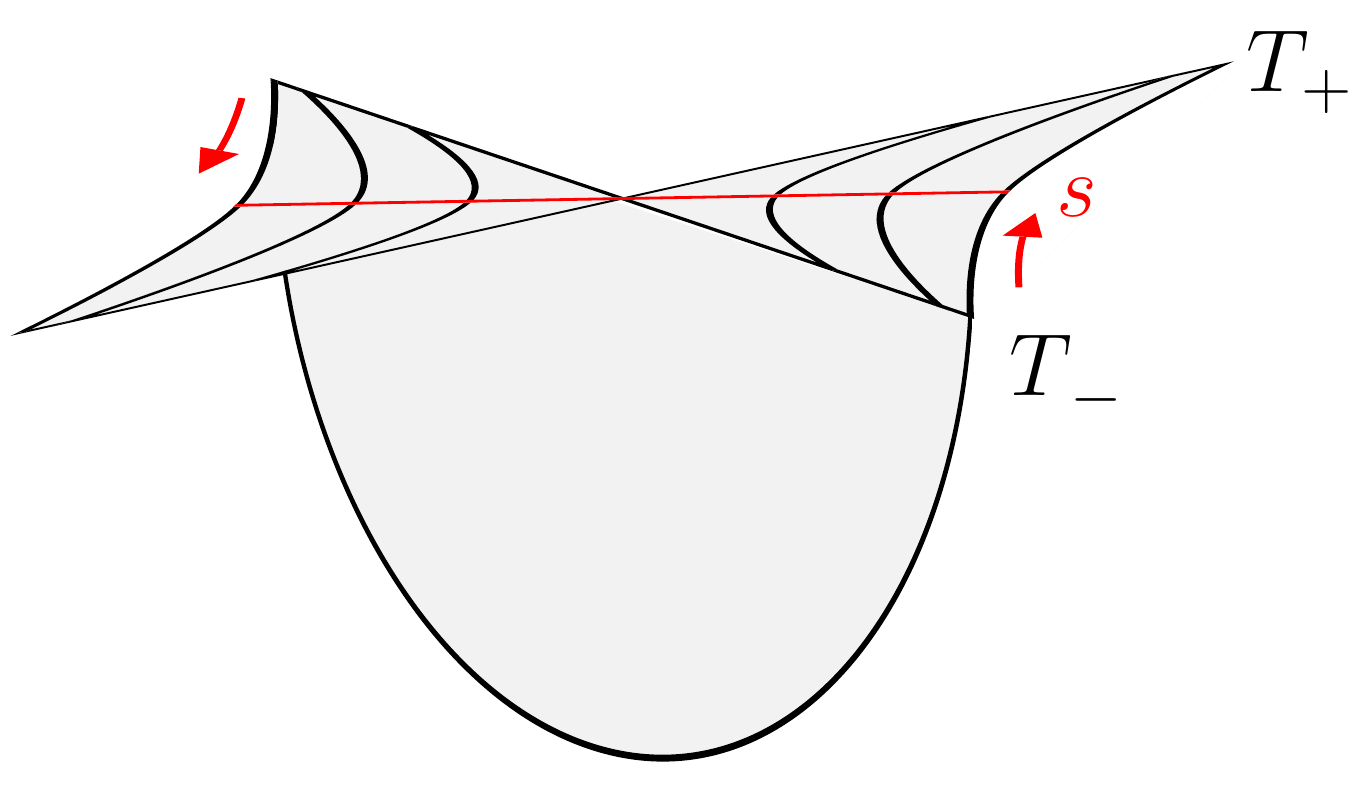}
\caption{
Upon building the Modular Hamiltonian via an Euclidean path integral, the picture shows its evolution on the L/R wedges given by the operator $\Delta^{is}$. }
\label{Fig:flow}
\end{figure}

Finally, it is worth noticing that in the Rindler spacetime,
the modular flow generated by $\Delta^{is}$ is naturally related to the sSK complexification of the evolution parameter in the extended geometry  of figure \ref{Fig:Camino}. In fact, let us consider the formula for a (initial) global state  \eqref{state-U}
\begin{equation}\label{Psi-U}
   |\Psi_\lambda\rangle \equiv  (U_\lambda(- i\pi R )\,\otimes\mathbb{I})|1\rangle\!\rangle 
\end{equation}
such that the  Modular Hamiltonian coincides with the generator of boosts and writes as \eqref{KR}. For accelerated observers it generates the time translations and the state lies on the surface corresponding to $T^-$ (see fig \ref{Fig:flow}); on the other hand \emph{the same} state, lying on the Cauchy surface with modular parameter\, $T^- + s$, also  obeys \eqref{Psi-U} and writes as 

\be 
|\Psi_\lambda(s)\rangle\equiv U_\lambda(s) \, U_\lambda(- i\pi R )\, U_\lambda(-s)|1\rangle\!\rangle ,
\ee
since as shown in  fig \ref{Fig:flow}, the operator $U_\lambda$ composes a boost rotation by $s$ of $\Sigma_R$ with a (Euclidean) rotation in $i \pi R$, and then with a boost by $-s$.
Using the \emph{tilde} conjugation and the TFD rules we can see that the modular evolution is realized by
\be 
|\Psi_\lambda(s)\rangle = U_\lambda(s) \, \tilde{U}_\lambda(s) \, U_\lambda(-i\pi R )\, |1\rangle\!\rangle = e^{is\, (\tilde{K}_\lambda-K_\lambda)}|\Psi_\lambda\rangle = \Delta^{is} |\Psi_\lambda\rangle\, , \Rightarrow \, \Delta\equiv e^{\tilde{K}_\lambda-K_\lambda} \,.
\ee

An important remark is that this modular flow is related to a complexified geometric flow in eq. \eqref{evols}, where  $\gamma^\mu (s \to \theta)$ and $\theta \in {\cal C}$ is the complex parameter of the closed symmetric SK path of figure \ref{Fig:Camino} which describes a closed curve that intersects $\Sigma_R(t)$ only in the point $X^\mu$. The union of all these curves covers completely the sSK geometry of Fig \ref{Fig:Camino}b: $W_{\cal C}$. In fact, the modular flow can evolve the local operators along these curves according to 
\be\label{evoltheta}
{\cal O}(r, X^i , t + \theta ) \equiv e^{-i\theta K_0} {\cal O}(X^\mu) e^{i\theta K_0}\;\;\;\;\;\theta \in {\cal C}\;, X^\mu\in \Sigma(t)
\ee
and defines a sort of extension of the operator algebra ${\cal A}(W_R)$, to local operators on all the points of  $W_{\cal C}$\,\footnote{On the other hand, one also have an extension associated to the commuting algebra $\tilde{{\cal A}}$.}. Then the Tomita-Takesaki theory might be interpreted as a constraint, relating these  operators ${\cal O}$   with certain operators $\tilde{{\cal O }}$ of the commuting algebra $\tilde{{\cal A }}$ when they act on a specific state. This point of view will become more precise in what follows.

\subsection{TFD formalism and the Tomita-Takesaki theory for holographic states}
\label{Sec:TomTakCFT}

The TFD formalism can be applied to study the entanglement in systems separated in two identical subsystems and reduced to one of them, such as the the vacuum of a QFT on a Rindler spacetime (or a black hole \cite{Peter 3, Peter 4}) reduced to a wedge $X^1\geq 0$.

According to this formalism, the condition that defines the (thermal) ground state and fixes the Bogoliubov transformation that relates this state with the (disentangled) vacuum of inertial observers \cite{UmezawaTFD,UmezawaTFD2}, is usually expressed as a constraint on the fields, ${\cal O} \in {\cal A}$ for an initial time
$t$ \cite{Peter 4}, 
\be \label{thermalstateconditionTFD-0} 
\,\left( \tilde{{\cal O}} (\tilde{X}, t)  -   {\cal O}^\dagger (X, t-\, i\beta/2 )\right)\;| \Psi_0\rangle =0\, , 
\ee
where $ \tilde{{\cal O}} (\tilde{X})\equiv  \widetilde{{\cal O}(X) }$ is defined by the tilde conjugation rule \cite{UmezawaTFD}, see App. \ref{tfd}.
In the Rindler space, this represents the field in the algebra $\widetilde{{\cal A}}$ of operators on the left wedge $W_L$\footnote{Formally, they are obtained by applying a CPT, composed with a particular rotation in a $\pi$ angle, on the fields ${\cal O}$, see \cite{Witten2018}}. 
This is known as the \emph{thermal state condition} in the TFD context \cite{UmezawaTFD,UmezawaTFD2}, and in particular has been studied in
spacetimes with event horizons, and interpreted as the quantum/operator formulation of the Unruh-trick \cite{Peter 4}, see discussion in \ref{TTexcited}. The geometric interpretation of this equation is that, in the state $|\Psi_0 (t)\rangle$, the field on the right can be related to the left ones by an analytically continued time evolution in $i\beta/2$.

Then, by identifying the tilde conjugation with the action of the operator $J$ of the polar decomposition \eqref{tomita}, and taking
\be\label{Delta-U0}
\Delta^{1/2} \equiv U_0(i\beta/2 ) \otimes  \tilde{U}_0(-i\beta/2 )  = e^{-\beta (K_0  - \tilde{K}_0)/2}
\ee
and $|\Psi_0 (t)\rangle$ given by \eqref{state-U} with $\lambda=0$ it can be verified that \eqref{thermalstateconditionTFD-0} is equivalent to the Tomita-Takesaki relation
\be \label{tomitat-condition}
S \, {\cal O} | \Psi_0\rangle =  {\cal O}^\dagger | \Psi_0\rangle 
\ee
which, being $| \Psi_0\rangle$ cyclic and separable (see App. \ref{tfd}),
guarantees that the operator $\Delta^{it} \equiv (S S^\dagger)^{it}$  defines the modular flow in the vacuum state $| \Psi_0\rangle$.
Therefore, we have shown with this simple example that in some cases, we can translate the Tomita-Takesaki theory to the TFD analysis, and interpret  eq. \eqref{tomitat-condition} as a constraint defining a state. 

Our main statement in this section is that the 
TFD constraint \eqref{thermalstateconditionTFD-0} can be generalized to excited states as \cite{Peter 4},
\be \label{thermalstateconditionTFD-lambda} 
\,\left( \tilde{{\cal O}} (\tilde{X}, t)  -   U_\lambda(-i\beta/2 ){\cal O}^\dagger (X, t )U_\lambda(i\beta/2)\right)\;| \Psi_\lambda\rangle =0\, . 
\ee
Thus, if we define $\Delta^{1/2}_\lambda$  by substituting $U_0 \to U_\lambda $ in \eqref{Delta-U0} it can be verified that
\be 
\label{tomitat-condition-lambdax} S_\lambda \, {\cal O} | \Psi_\lambda\rangle =   {\cal O}^\dagger | \Psi_\lambda\rangle,
\ee
with
\be 
S_\lambda\equiv J \Delta_\lambda^{1/2}\;,\qquad\qquad \Delta_\lambda=\rho^{-1}_\lambda\otimes\rho_\lambda.
\ee
Notice that $J$ remains unchanged under the deformation. This follows from the fact that the deformation can be equivalently created by operators acting only on either side of the d.o.f splitting, see \eqref{state-U}. A more explicit derivation of this result is presented in Sec. \ref{Tomita-Bulk}. This type of deformations are known to preserve the $J$ operator and are contained in whats called the ``standard cone'', see \cite{Haag}. From a TFD perspective, the formalism naturally admits excited states, see \cite{Takashi,Thermal-Coherent} for example, without deforming the tilde map between the duplicated theories. 

Let us see, for instance, that this constraint is trivially satisfied for the (holographic) excited states constructed with a time-independent source $\lambda\equiv \lambda(X)$, in an arbitrary QFT.
In this case the Modular Hamiltonian writes, see \eqref{KR},
\be\label{H-deformed-tindepedent}
K_0 \to K_\lambda \equiv K_0 + \int_{\Sigma_R} dX^{d-1}\, \lambda(X) \, {\cal{O}}(X)\,.
\ee
and, because of the Bisognano- Wichmann theorem\cite{Bisognano}, the time evolution coincides with the modular evolution generated by this operator. Consequently, the state defined as 
\be
| \Psi_\lambda \rangle  = U_\lambda(-i\beta/2) | 1\rangle\! \rangle = e^{-\beta K_\lambda /2} | 1\rangle\! \rangle
\ee
is the new TFD-vacuum for the deformed Hamiltonian \eqref{H-deformed-tindepedent}, although it is an excited state of the original (undeformed) theory. In other words, the constraint \eqref{thermalstateconditionTFD-lambda} reduces to \eqref{thermalstateconditionTFD-0}.

We would like to close this Section by pointing out that in CFT, this construction can be straightforwardly extended to regions bounded by a sphere through the CHM map, or regions conformally related to a Rindler wedge.

We emphasize that the computations done so far in this section are on QFT and hold for arbitrary coupling constant and $N$. In this sense our results are non-perturbative. However, when $\lambda$ depends on $\tau$ it is difficult to prove that the equation \eqref{thermalstateconditionTFD-lambda} is satisfied by the objects identified in \eqref{tomitat-condition-lambdax}.
There are two scenarios where this can be done explicitly. In the first case one may consider a perturbative expansion of the QFT to a fixed order in the coupling constant, thus proving the statement in a weakly coupled regime. The second set-up would be that of a strongly coupled (large $N$) CFT such that a holographic description is available. This again allows an expansion for single trace CFT operators, but this time in terms of the bulk ladder operators, see \cite{BDHM,BDHM2}. In the next section we will follow the latter approach and compute the dual Modular Hamiltonian to $K_\lambda$ in the bulk in the large $N$ limit, where we will be able to provide more explicit results.


\section{The gravity dual of the Modular Hamiltonians (at large $N$)} 
\label{gravity}

In the previous section we were computing the Modular Hamiltonian and modular flows for the particular excited states introduced in Section 2 from a QFT point of view. We concluded that the result can be written as \eqref{density-rindler} for an equipartite subsystem and \eqref{guessKD} for the spherical entangling surface. But, we can't ensure that this result is the one related to the modular operator $\Delta$ of the TT theory in the general case with explicit $\tau$-dependence of the source $\lambda$. In the present section we will use the fact that we know the precise holographic dual of the  state \eqref{density-rindler} and of the condition \eqref{thermalstateconditionTFD-0} to show that it satisfies the TT constraint in the bulk. Then, we will see in this way that the Modular Hamiltonian computed in the previous section in the context of a CFT is the one associated with $\Delta$ in the strong coupling limit of the field theory.
We will also provide explicit examples of the modular flow induced by our excited Modular Hamiltonians on both sides of the duality. We will conclude the section mentioning how we can explicitly compute matrix elements of Modular Hamiltonians for the excited states using the JLMS proposal.

\subsection{Expected results}

The cases that we have studied can be considered dual to spacetimes with a Killing  vector in the bulk: $\zeta^\mu = \partial^\mu/\partial t$, such as AdS-Black Holes, Rindler-AdS, or isometric mappings of these spaces (e.g. regions whose asymptotic boundaries are ball-shaped \cite{Casini2011,Faulkner:2013ica}). In these cases, $\zeta$ is bifurcating on the entangling surface, given by the RT recipe (see \cite{Casini2011}).
 Therefore, according to the prescription 
 discussed in section \ref{modularqft}, the expectation is that the (bulk) modular flow shall also be determined by the euclidean evolution operator $U_{bulk}(0, i2\pi)$, but the $\lambda$-deformations shall be described as no vanishing Dirichlet asymptotic BCs on the bulk fields \cite{Peter 1, Peter 2, Peter 4}.
 
 In the following analysis of the bulk theory we will assume a large N approximation, which in particular, supposes non-backreacting and weakly coupled QFT in the gravity side, i.e. the existence of ladder operators. For simplicity we will also consider a real scalar field $\Phi$.

\begin{figure}[t]\centering
\includegraphics[width=.49\linewidth] {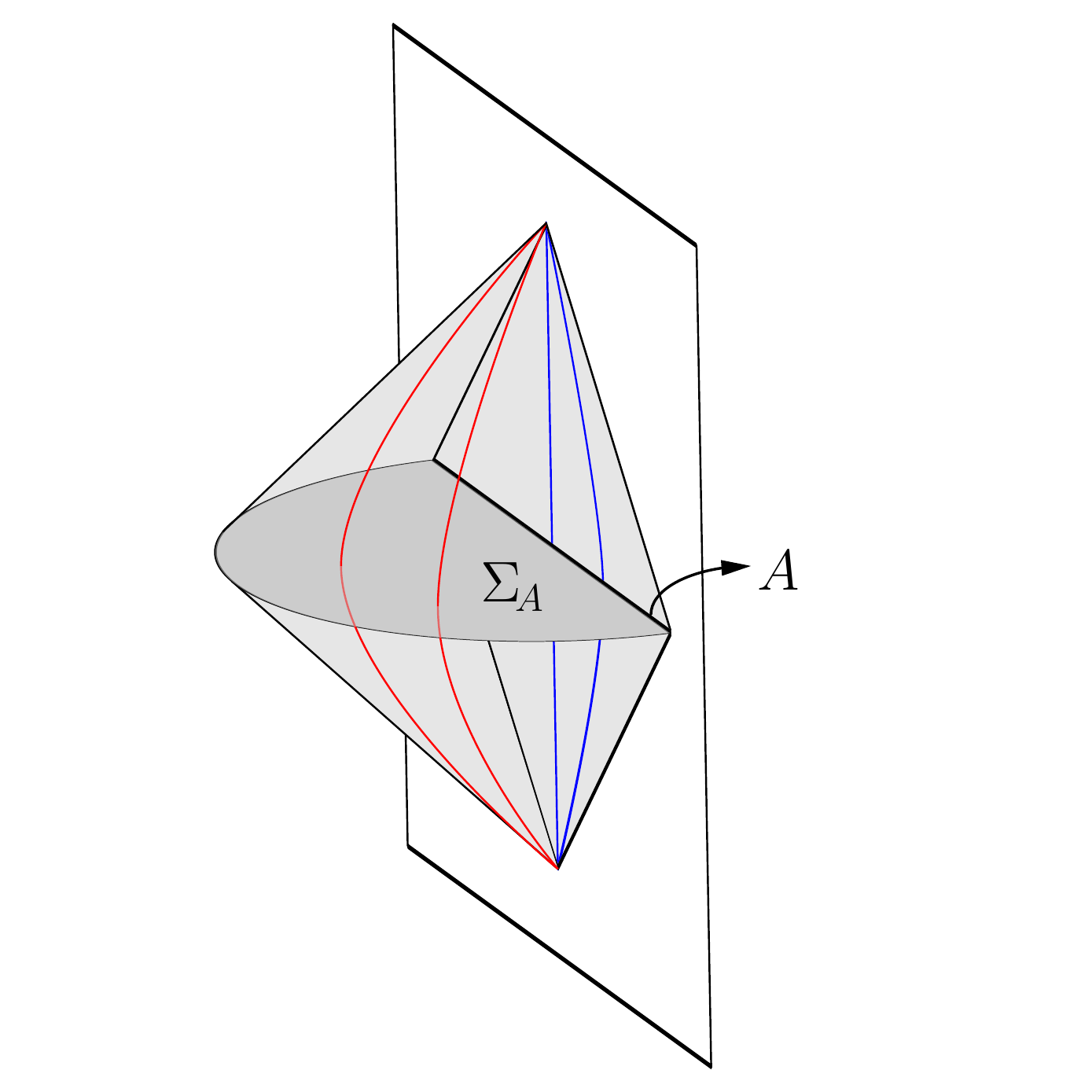}
\caption{The figure shows a subregion $A$ of a system defined on the white plane and its causal diamond, as well as its entanglement wedge inside its holographic dual. Assuming a bulk timelike killing vector $\zeta$, whose flows are shown in red, one can define the spacelike surfaces $\Sigma_A$. The blue lines are the flows that live on the boundary.}
\label{Fig:FlowRindler}
\end{figure}

The canonical Hamiltonian $H=H(\Phi, \Pi)$, derived from the action $S[\Phi]$, is the operator that generates the $t$-translations on the entanglement wedge (see red lines in fig. \ref{Fig:FlowRindler} ) . Thus, for the vacuum state $\lambda=0$
 one can write \eqref{density-bulk} as 
\be\label{density-bulk-H}
 \rho_0 [\phi^+, \phi^-]\equiv \langle \phi^+ |e^{-2\pi H} | \phi^-\rangle\;,
 \ee
 and so the \emph{Modular Hamiltonian} in this case is simply $K_0 \equiv 2\pi H$, and the modular flow will be defined by
 \be
 U(t)= e^{-2\pi\,i t\, H}\,.
 \ee
The simple form of the holographic formula \eqref{density-bulk} suggests an ansatz for holographic excited states defined by non trivial asymptotic conditions on the Euclidean sections of the spacetime. Thus, we expect that at large $N$, the backreaction is negligible and the Killing vector remains $\zeta^\mu$, and thus  $K_{\lambda}$ will depend on $\lambda$ in the following way
  \be\label{Hlambda}  K_\lambda \propto H[\phi+ f_\lambda\,, \pi - \partial_t f_\lambda ] \ee
 where $f_\lambda $ is a particular solution of the classical e.o.m. satisfying non vanishing asymptotic boundary conditions $(\lambda ; \lambda^\star) = f_\lambda |_{\partial {\cal{E}}}$  on the euclidean pieces of the spacetime. 
 This is nothing but the canonical Hamiltonian under the substitution $\phi \to \phi_\lambda \equiv \phi+ f_\lambda $ in the action.

Although the explicit computation will be done later, a simple path integral argument for it is that this field redefinition transforms the expression \eqref{density-bulk} to
 
 \be\label{density-bulk-redef}
 \rho_\lambda [\phi^+, \phi^-]\equiv \langle \phi^+ |\rho_\lambda | \phi^-\rangle \equiv \langle \phi^+ |{\cal P} e^{-\int^{2\pi}_0\, d\tau H_\lambda} | \phi^-\rangle\;= \int_{\lambda=0, \phi^\pm} [D\Phi] \; e^{-S[\Phi\, + \, f_\lambda]}\;,
 \ee
 where the sum is over fields $\Phi$ with vanishing asymptotic b.cs. So this can be thought as a new theory that under certain conditions, in particular at large $N$, does not break the time-translational symmetry and the Hamiltonian canonically derived from the action $S[\Phi\, + \,f_\lambda]$  coincides with \eqref{Hlambda}. Furthermore, it would generate the modular/dynamic flow in the bulk, and \eqref{density-bulk-redef} can be expressed as 
  \be\label{density-bulk-H-lambda}
 \rho_\lambda \,=\, e^{-2\pi H_\lambda}
 \ee
 Thus, one can show this anzatz by interpreting the above substitution as a \emph{canonical transformation} of the fields (and its canonically conjugated momenta), 
  one can see that the equations of motion
 \be\label{eomHeisemberg} \partial_t \phi_\lambda = i [H_\lambda , \phi_\lambda] \;\;,\,\,\,\partial_t \pi_\lambda = i [H_\lambda , \pi_\lambda] \ee 
 are preserved, as the fields are promoted to operators according to the rule 
\be\label{CCR} [ \phi_\lambda (x_1) , \pi_\lambda (x_2) ] = i\,\delta_{1,2} \qquad\;\; x_{1,2} \in \Sigma .\ee

Then, the quantization of this theory in the Heisenberg picture consists in finding the general solution to the equations \eqref{eomHeisemberg} (while the state keep the same) provided \eqref{CCR}. Assuming that 
 $\phi$ is the most general solution of the problem with vanishing (asymptotic) boundary conditions, e.g. at large-N approximation, it is a linear combination of the normalizable modes.
Therefore, $\phi_\lambda , \pi_\lambda $ is nothing but the most general field (solution) of the equations of motion, and the condition \eqref{CCR} is automatically satisfied, by demanding that the particular solution $f_\lambda$ be a c-number. 
 
Then, the time-evolution for any operator $A(\phi_\lambda , \pi_\lambda)$ of the theory is given by
\be
 A(t) = e^{i t H_\lambda} A(0) e^{-i t H_\lambda} 
 \ee
 and in particular 
 \be\label{modular-phi} 
 \phi(t) = e^{i t H_\lambda} \phi(0) e^{-i t H_\lambda} - \left(f_\lambda(x,t) - f_\lambda(x,0)\right)
 \ee
 \be\label{modular-pi} 
 \pi(t) = e^{i t H_\lambda} \pi(0) e^{-i t H_\lambda} - \left(\dot{f}_\lambda(x,t) - \dot{f}_\lambda(x,0)\right)
 \ee
 Then \eqref{Hlambda} is the Modular Hamiltonian corresponding to the excited (holographic) state $|\Psi_\lambda\rangle$, since it generates the modular flow. We will see in Sec. \ref{Tomita-Bulk} that  by virtue of large-$N$ approximation, the field equations can be assumed to be linear, and then one can exactly consider the canonical quantization. 
 
\subsection{Tomita - Takesaki formalism in the bulk}\label{Tomita-Bulk}

Here we will give the proof in the bulk of the formulae derived in Sec. \ref{Sec:TomTakCFT} for the Modular Hamiltonians for the excited states. The central point of the argument resides again in the relation between TFD and Tomita-Takesaki theory. Starting from the TFD vacuum thermal equilibrium condition, we will show explicitly how to map it to the Tomita-Takesaki equation and how to extract the exact $\Delta_0$ and $J$. From there, we will deform the original thermal equilibrium condition to include the holographic excited states but we will still be able to perform the mapping to the Tomita-Takesaki equation. Thus, we will be able to identify the excited $\Delta_\lambda$ and $J$ for the bulk theory. 

\subsubsection{From TFD to Tomita-Takesaki: Vacuum State}

We will extract from the thermal state condition of the TFD vacuum state both the modular and $J$ operators. We start from the bulk analog of \eqref{thermalstateconditionTFD-0},
\begin{equation}\label{vinculo0}
    \left( \Phi_R(t) - \Phi_L(t-i\beta/2) \right)|\Psi_0\rangle\ = \left( \Phi_R(t) - U_0 (-i\pi, 0)\Phi_L(t) U_0 (0, i\pi)  \right)|\Psi_0\rangle\ = 0,
\end{equation}
and show that this can be rewritten as (cf. with \eqref{Stt})
\begin{equation}
    S \Phi_R(t)|\Psi_0\rangle = \Phi^\dagger_R(t) |\Psi_0\rangle= \Phi_R(t) |\Psi_0\rangle,\label{TomitaT0}
\end{equation}
where we have used the fact that the fields $\Phi$ are Hermitian. 
Recall that the TFD formalism readily provides an antiunitary tilde operation which map the operators from R to L and vice-versa, $\tilde \Phi_R(t)=\Phi_L(t)$ and $\tilde \Phi_L(t)=\Phi_R(t)$. This can be represented as an operator $J=J^{-1}$ such that $\tilde \Phi_R(t) \equiv J \Phi_R(t) J^{-1} = \Phi_L(t)$. Notice that $J$ does not factorize into L and R pieces. The specific form of $J$ is however not important for our purposes but the interested reader can see \cite{Witten2018} for details.

The central piece of this argument is the fact that one can build the TFD vacuum out of an operator with support only on one of the sides, i.e. $U_0(0,-i\pi)$, on the identity operator $|1\rangle\!\rangle$ introduced before, see Appendix \ref{tfd},
\begin{equation}\label{Psi0}
    |\Psi_0\rangle\ \equiv U_0(0,-i\pi)\otimes\mathbb{I}|1\rangle\!\rangle = \mathbb{I} \otimes U_0(0,-i\pi) |1\rangle\!\rangle\;.
\end{equation}
This is a consequence of the Reeh-Schlieder theorem \cite{Reeh}, and the highly entangled nature of the vacuum, manifestly expressed in the  state $|1\rangle\!\rangle$.

We now demonstrate the connection between \eqref{vinculo0} and \eqref{TomitaT0} starting from a more explicit version of \eqref{vinculo0}
\begin{align}\label{demon0}
     \mathbb{I} \otimes \Phi_R  |\Psi_0\rangle\ &= ( U_0(-i\pi,0)\Phi_L U_0(0,i\pi)  ) \otimes \mathbb{I} |\Psi_0\rangle\ \\
     &= ( U_0(-i\pi,0) \Phi_L  U_0(0,i\pi) \otimes \mathbb{I} )  ( \mathbb{I} \otimes   U_0(0,i\pi) U_0(-i\pi,0) ) |\Psi_0\rangle\  \nn\\
     &= (U_0(-i\pi,0)\otimes U_0(0,i\pi)) \;  (\Phi_L  \otimes \mathbb{I} ) \;(U_0(0,i\pi) \otimes   U_0(-i\pi,0) ) |\Psi_0\rangle\ \nn \\
    &= (U_0(-i\pi,0) \otimes U_0(0,i\pi))  \; J\; (\mathbb{I} \otimes \Phi_R )\;J \;(U_0(0,i\pi) \otimes U_0(-i\pi,0) ) |\Psi_0\rangle\  \nn\\
    &\equiv \Delta^{-\frac 12} J (\mathbb{I} \otimes \Phi_R )J \Delta^{\frac 12} |\Psi_0\rangle\ \nn\\
    &= S (\mathbb{I} \otimes \Phi_R ) S |\Psi_0\rangle\ =S (\mathbb{I} \otimes \Phi_R ) |\Psi_0\rangle,\nn
\end{align}
where in the second line we inserted $\mathbb{I}=U_0(0,i\pi) U_0(-i\pi,0)$ and in the last equality we used $S |\Psi_0\rangle=J \Delta^{\frac 12} |\Psi_0\rangle=|\Psi_0\rangle$ which is trivial if $S$ is the correct Tomita operator, but from our perspective this is still left to prove. Actually, in order to meet the Tomita-Takesaki theorem conditions, we need to prove both $\Delta^{\frac 12} |\Psi_0\rangle = J|\Psi_0\rangle = |\Psi_0\rangle$ independently. The condition on $J$ follows trivially from the fact that $U_0(-i\pi,0)=e^{-\frac \beta 2 H}$, where $H$ is an hermitian Hamiltonian which can be diagonalized with real eigenfunctions and the fact that it acts trivially on $|1\rangle\!\rangle$ by definition. The demonstration then follows as
\begin{equation}
    J|\Psi_0\rangle = J U_0(0,-i\pi)\otimes\mathbb{I}|1\rangle\!\rangle = \left[J ( U_0(0,-i\pi)\otimes\mathbb{I} ) J \right] J |1\rangle\!\rangle =\mathbb{I} \otimes U_0(0,-i\pi)  |1\rangle\!\rangle= |\Psi_0\rangle\;.
\end{equation}
 Notice that the operator $J$ is antiunitary, but $U_0(0,-i\pi)\in\mathbb{R}$ is a Wick rotation an analytical extension of a unitary evolution operator. It is also immediate to show
\begin{equation}
    \Delta^{\frac 12}|\Psi_0\rangle = (U_0(0,i\pi) \otimes U_0(-i\pi,0)) (U_0(-i\pi,0)\otimes\mathbb{I})|1\rangle\!\rangle = \mathbb{I} \otimes U_0(0,-i\pi)  |1\rangle\!\rangle= |\Psi_0\rangle,
\end{equation}
which completes the demonstration. Note that \eqref{Psi0} was crucial.

\subsubsection{TFD to Tomita-Takesaki: Excited States}

Once proven for the vacuum, we will consider the holographic excited states,
\begin{equation}\label{Psi1}
    |\Psi_\lambda\rangle \equiv U_\lambda(0,-i\pi) \otimes \mathbb{I} |1\rangle\!\rangle = \mathbb{I} \otimes U_\lambda(0,-i\pi) |1\rangle\!\rangle \;.
\end{equation}
Note that the state admits two equivalent ways of defining it via operators acting only on either L or R. One could readily argue that the $J$ operation should not be deformed, see discussion below \eqref{tomitat-condition-lambdax}.

The constraint on the excited state is
\begin{equation}\label{vinculo1}
     \left[ \Phi_R(t) - U_\lambda(-i\pi,0) \Phi_L(t) U_\lambda(0,i\pi) \right]|\Psi_\lambda\rangle = 0,
\end{equation}
which will lead to
\begin{equation}\label{TomitaT1}
    S_\lambda \Phi_R(t)|\Psi_\lambda\rangle = \Phi^\dagger_R(t) |\Psi_\lambda\rangle= \Phi_R(t) |\Psi_\lambda\rangle\;,\qquad\qquad S_\lambda\equiv  J \Delta_\lambda^{\frac 1 2}=\Delta_\lambda^{-\frac 1 2} J
\end{equation}
where we have used again hermiticity of the fields $\Phi$.

The demonstration of \eqref{TomitaT1} from \eqref{vinculo1} follows as in \eqref{demon0}.
As before, one has to prove that 
$$S_\lambda |\Psi_\lambda\rangle = J |\Psi_\lambda\rangle= \Delta_{\lambda}^{\frac 12} |\Psi_\lambda\rangle= |\Psi_\lambda\rangle\;,$$
which also follow analogously from the vacuum computation. This completes the demonstration.

As a summary, the main result is that for the set of excited states and systems described above, we get a closed expression for the Modular operator, which can be written as
\begin{equation}\label{ModOperatorResult}
    \Delta_\lambda = \rho_\lambda^{-1} \otimes \rho_\lambda \;,\qquad  \langle \phi_+|\rho_\lambda|\phi_-\rangle = \int {\cal D}\Phi_{\lambda;\phi_\pm} \;e^{-S_E[\Phi]}\;. 
\end{equation}
Notice again that the theory is undeformed, and only the boundary conditions are affected. From here we can compute a reduced Modular Hamiltonian, which coincides with the one found in \cite{Van17} and \cite{Vanraam2019} for these type of states at first order in in $1/N$.


To conclude this subsection we observe that, once \eqref{ModOperatorResult} is known, one can also obtain the modular operator for any other (that can be non equipartite) subsystems connected to \eqref{ModOperatorResult} via an isometry of AdS. Since we are considering an excited state, both the region and the state are affected by the transformation. As the excited states are created by a perturbation localized only on one the subsystems, the AdS isometries will still map the deformation inside of the transformed subsystem.

\subsection{Computing the bulk Modular Hamiltonian and bulk modular flow at large $N$}\label{modbulk}

Consider an equipartite bulk spacetime with a Killing vector $\zeta^\mu$, which is bifurcating on the entangling  surface, and it is holographically dual to the Rindler spacetime studied in the previous Sections. We can describe this with the following metric:
\be\label{adsR}
ds^2=-u^2dt^2+\frac{du^2}{1+ u^2}+\left(1+u^2\right)\,\left(d\chi^2 + \sinh^2\chi\,d\Omega^2_{d-2}\right)
\ee
where the coordinate $y \equiv (\chi, \Omega_{d-2})$ involves a non-compact component $\chi$, while $\Omega_{d-2}$ describes a $(d-2)$-sphere. The holographic coordinate $u$ can be extended to take all the real values (e.g. \cite{BootsBrazil2}); thus  $u>0$ stands for the wedge that, after a suitable change of coordinates: $u^2 \to u^2 - 1$, is dual to the  a hyperbolic cylinder on the boundary \cite{Faulkner:2013ica}, that can be conformally mapped to a ball shaped region, or to one of the two wedges of the flat Rindler spacetime \cite{Casini2011}. Figure \ref{Fig:FlowRindler} illustrates how the Killing vector $\zeta \equiv \partial_t$, asymptotically coincides (up to a conformal map) with the vector $\partial_t$ of the boundary metric 
\eqref{RindlerCFT}. The sSK extension of this geometry is similar to Fig. \ref{Fig:Camino}, but the theory here is to be sourced by a Dirichlet BC at the asymptotic boundary of the euclidean regions $\partial {\cal E}^\pm$, see Fig. \ref{fig:evo2}(b).

Consider a canonically quantized free scalar field $\Phi$ in the bulk. This is essentially the behaviour of all the fields of the bulk theory in the large $N$ approximation. The general solution on the entanglement wedge (one of the two sides of the bulk spacetime, say $u>0$) writes 
\be\label{quantized-field}
\Phi (u,y,t) = \sum_{n} \; a^\dagger_n \phi_n(u, y ,t) + h.c. 
\ee
The eigenfunctions $\phi_n(u, y ,t)$ are assumed to be an orthonormal basis of the space of (positive energy) solutions of the e.o.m., and the subindex $n$ collectively denote its quantum numbers.

The global state in the bulk theory, can be computed through the formula
\be
|\Psi_\lambda\rangle = U_\lambda (0,- i\pi) |1\rangle\!\rangle
\ee
where $U_\lambda (0,-i\pi)$ is the (euclidean) evolution operator in the Schr\"odinger picture. A convenient trick is to transform this to the Interaction Picture, in which the state can be expressed as
\be\label{psi-desplazamiento}
|\Psi_\lambda\rangle = D(\lambda) |\Psi_0\rangle = D(\lambda) e^{-\pi H}|1\rangle\!\rangle,
\ee
where $D(\lambda) = \prod_n D(\lambda_n)$ is the (unitary) \emph{displacement operator} such that $D(\lambda_n) a_n D^\dagger(\lambda_n) =a_n + \lambda_n$.
Then using \eqref{densitystate} we get 
\be
\rho_\lambda = D(\lambda) \rho_0 D^\dagger(\lambda) = D(\lambda) e^{-2\pi H} D^\dagger(\lambda),
\ee
which is nothing but a thermal coherent state. By expressing the Hamiltonian as $K_0 \equiv 2\pi H= 2\pi\sum_n \; w_n : a_n \,a_n^\dagger\,: $ and certain algebraic work using the BCH formulas one obtains
\be
\rho_\lambda = e^{-2\pi H_\lambda},
\ee
where
\be  H_\lambda = D(\lambda) H D^\dagger(\lambda) = \sum_n w_n D(\lambda_n) \,: a_n \,a_n^\dagger\,: \,D^\dagger(\lambda) =\sum_n \; w_n : ( a_n + \lambda_n) (a_n^\dagger + \lambda_n^*):\;\;.
\ee
Here we stand for $\lambda$ the decomposition of the source in (euclidean) normal modes \cite{Peter 1, Peter 4}
\be\label{lambda-n}\lambda_n  \equiv \lim_{|u|\to \infty} u^{\Delta}\;\int_{\Sigma_R} dy\, \int_0^\pi \, d\tau \,\lambda (y, \tau) \, \phi_n (u , y, -i \tau) \,\, ,  \ee

This is the expected expression \eqref{Hlambda} in terms of the frequency components of the fields and momenta, and the displacement operator realizes the canonical transformation in these variables. In fact, one of the results of the present analysis is that, at large $N$, the holographic excitations consist of a family of canonical transformations, parameterized by the holographic source $\lambda(y,\tau)$.
It is worth emphasizing that here, the source $\lambda$ can depend arbitrarily on the coordinates of the half euclidean boundary $\partial {\cal E}^-$.

Note that the map 
\be H\to H_\lambda \qquad a_\lambda \equiv a + \lambda \qquad a^\dagger_\lambda \equiv a^\dagger + \lambda^*, \ee
is a canonical transformation. 
The label $n$ for each normal frequency mode from \eqref{lambda-n} is left implicit.
One can verify that the canonical commutation relations are preserved for the new set of ladder operators and therefore, the e.o.m for the Heisenberg operators are
\be\label{eomHeisemberg-ladder} 
\dot{a}_\lambda = [a_\lambda \,,\, H_\lambda ] =\, w \, a_\lambda \;,\qquad \dot{a}^\dagger_\lambda =[a^\dagger_\lambda \,,\, H_\lambda ] =\,- w \, a^\dagger_\lambda\;. \ee
Here we stand for $\lambda$ the decomposition of the source in (euclidean) normal modes, eq. \eqref{lambda-n}.
Since the parameter of the evolution generated by $H_\lambda$ is often called $s$,  from \eqref{eomHeisemberg-ladder} one gets the equations
of evolution
\be\label{eom-lambda} i \frac{da_\lambda}{ds}  =\, w \, a_\lambda \;,\qquad i \frac{da_\lambda^\dagger}{ds} =\,- w \, a^\dagger_\lambda\;, \ee
which can be integrated to obtain the explicit modular evolution
\be\label{a(s)}
a_\lambda (s) \equiv e^{-i sH_\lambda}\, (a_\lambda) e^{i sH_\lambda}= e^{isw} a_\lambda \qquad a^\dagger_\lambda (s) \equiv e^{-i sH_\lambda}\, (a_\lambda^\dagger) e^{i sH_\lambda}= e^{-isw} a^\dagger_\lambda\;.
\ee

Define the (deformed) field operator as
\begin{align}\label{evol-phi-lambda}  \Phi_\lambda(u,y,0) &\equiv D(\lambda) \Phi(u,y,0) D^\dagger(\lambda) \\&= \sum_{n}  \;\; D(\lambda_n) \, a_n^\dagger \, \,D^\dagger(\lambda_n) \phi_n(u, y ,t=0)   +  h.c.\;\\
&= \sum_{n}  \;\; (a_n^\dagger + \lambda_n^*)\; \phi_n(u, y ,t=0)   +  h.c.\;\\
&= \sum_{n} \; (a_\lambda^\dagger)_n \;\phi_n(u, y , 0) + h.c., \end{align} 
therefore, we can compute the $s$-evolution of this operator

\be  \rho_\lambda^{is} \Phi_\lambda (u,y,0) \rho_\lambda^{-is}\;= 
\sum_{n} \; e^{-i sH_\lambda}\, (a_\lambda^\dagger)_n e^{i s H_\lambda}\,\phi_n(u, y ,t \to s) + h.c.\,,
\ee
that by virtue of \eqref{a(s)}, takes the form:
\be\label{evol-phi-lambda2}
\sum_{n} \; \, (a_\lambda^\dagger)_n \, e^{-i s w_n}\,\phi_n(u, y , 0) + h.c.
=\sum_{n} \;  (a_n^\dagger + \lambda_n^*) \phi_n(u, y , s) \,+ \,c.c.
=\Phi(u,y,s) +  f_\lambda (u,y,s),
\ee
where
\be
\Phi(u,y,s) \equiv \sum_{n} \; a_n^\dagger \, \phi_n(u, y , s) \,+ \,h.c. \qquad    f_\lambda (u,y,s) \equiv \sum_{n} \;   \lambda_n^*\phi_n(u, y , s) \,+ \,c.c.
\ee
$\Phi(u,y,s)$ is the canonically quantized field as the time coordinate $t$ is interpreted as the parameter $s$, and $f_\lambda(u,y,s)$ is the solution of the classical e.o.m. on the entanglement wedge, with asymptotic boundary conditions $(\lambda , \lambda^\star)$ on $\partial {\cal E}$ and $0$ otherwise (see Fig \ref{Fig:Camino}). 

On the other hand, from \eqref{evol-phi-lambda} notice that
\be  \Phi_\lambda(u,y,0) = \Phi(u,y,0) + f_\lambda (u,y,0)  \;,
\ee
thus,
\be  e^{-i sH_\lambda} \,\Phi_\lambda(u,y,0) \, e^{i sH_\lambda} = e^{-i sH_\lambda} \Phi(u,y,0) e^{i sH_\lambda} + f_\lambda (u,y,0) \;.
\ee
Comparing finally with \eqref{evol-phi-lambda2}, we obtain the modular evolution of the original field
\begin{align}\label{bulkmod}
e^{-i sH_\lambda} \Phi(u,y,0) e^{i sH_\lambda} &= \Phi(u,y,s) + f_\lambda (u,y,s) - f_\lambda (u,y,0). \\
&= \Phi(\gamma^\mu(s)) + f_\lambda (\gamma^\mu(s)) - f_\lambda (\gamma^\mu(0))
\end{align}
where $\dot{\gamma}^\mu(s) \equiv \tau^\mu$ is the timelike Killing vector of the AdS-Rindler spacetime. In these coordinates $\gamma^\mu(s) = (u, y,  t + 2\pi s)$. This result resembles the one obtained in \cite{Liu} in the axiomatic Quantum Field Theory context on flat spacetime, but the non trivial fact here is that, because of holography, $f_\lambda(u,y,s)$ is the (unique) classical solution to the boundary problem schematically described in Fig. \ref{Fig:Camino}.

Finally, it is worth emphasizing that by virtue of the BDHM prescription \cite{BDHM}:
\be\label{BDHMaa}
{\cal O}(y, 0) = \lim_{u\to \infty } |u|^\Delta \; \Phi (u, y, 0)
\ee
and by assuming the holographic duality between the (QFT/bulk) modular flows, one can compute
the modular evolution of operators ${\cal O}(s) \in {\cal A}$ in the dual CFT by
\be\label{conjecture}
{\cal O}(y, s) = \lim_{u\to \infty }\, |u|^\Delta\, e^{-i s K_\lambda} \Phi (u, y, 0) e^{i s K_\lambda}\, ,
\ee
that in the case studied here (at the large $N$) gives:
\be\label{boundarymod}
{\cal O}(y, s) = \lim_{u\to \infty }\,|u|^\Delta\; \Phi(u,y,s) +  \lim_{u\to \infty }\,|u|^\Delta\; \left[ f_\lambda (u,y,s) - f_\lambda (u,y,0) \right]\;.
\ee
Notice that the last terms does not vanish in the $u\to \infty$ limit. 
This (radial) limit generates a set of operators that are included in the boundary algebra ${\cal A}$.

We would like to conjecture that the formula \eqref{conjecture}, to compute the modular evolution of operators in a holographic field theory from its gravity dual, has general validity for arbitrary regions $A$ and states (see discussion of Sec. \ref{Sec:AdS-CFT}).

\subsubsection{ An explict example of  Modular Flow and Tomita-Takesaki theory in AdS$_{2+1}$/CFT$_{1+1}$}\label{TTexcited}

This section is devoted to  solve
the normal modes of a scalar field in AdS$_{2+1}$-Rindler exactly, and compute the modular flow expressed in eq \eqref{boundarymod} for some particular examples of excited states. Moreover, we will show the realization in this explicit example of the Tomita-Takesaki construction and its relation with the constraints \eqref{vinculo0} and \eqref{vinculo1} for an excited state. The results and remarks achieved also hold for the extended BTZ spacetime.  

Consider a free scalar field $\Phi$ in AdS$_{2+1}$ in Rindler coordinates, that explicitly split the system in two equal halves
\be\label{metric-example}
ds^2=-u^2dt^2+\frac{du^2}{1+ u^2}+\left(1+u^2\right)\,d\chi^2
\;,\qquad\qquad (\square-m^2)\Phi(u,\chi,t)=0.\ee

Observe that if the coordinate is extended to cover all the real interval $u\in (-\infty , \infty)$ this metric captures both sides L ($u\leq 0$) and R ($u \geq 0$),
and the boundary of the subsystem $R/L$ is placed on the Killing horizon $u=0$, such as in the Section above. Clearly, the sSK extension of this geometry is similar to Fig. \ref{Fig:Camino}, and the ground states and holographic excitations correspond to the euclidean pieces as explained in the paper.

Recalling from the previous Section that the quantized fields on the left wedge ($u<0$) can be constructed from the right ones $\Phi$ on the right (entanglement) wedge by the operation $J$, or equivalently the tilde conjugation rules \eqref{Tilde-rules} of the TFD formalism, we can express the global solution on the Lorentzian regions as
 \be 
\Phi(u, \chi, t) = \Tilde{\Phi} (u, \chi, t)\; \Theta(-|u|) + \Phi (u, \chi, t)\; \Theta(|u|).
\ee
where $\Theta(x)$ is the Heavyside step function. Notice that $\partial / \partial t$ is a Killing vector, and $\Phi$ (and ${\tilde\Phi}$) can be canonically quantized in terms of (positive-energy) normalizable modes as in equation \eqref{quantized-field} }.

In this metric, the normal modes $\phi_n$ in \eqref{quantized-field} form a continuous basis of eigenfunctions (with $n\equiv (\omega, l) $), so the field can be written as
\begin{equation}\label{g}
\Phi (u, \chi, t)=\int_{\omega>0} \!\!\!\!\!\!d\omega dl \;a_{\omega l} \phi_{\omega l}(u,\chi,t) + hc,\;\;\qquad\phi_{\omega l}(u,\chi,t)=  {\cal N}\!\!_{\omega l} \; e^{-i \omega t+i l \chi} \left[f_{\omega l}(u)-f_{-\omega l}(u)\right] \;\;,\;\;\; u\geq 0
\end{equation}
\begin{equation}\label{f}
f(\omega,l,u)\equiv 
{\cal C}_{\omega l\Delta} \; 
r^{-\Delta } \left(1-\frac{1}{u^2}\right)^{i\frac{\omega }{2}} \, _2F_1\left(\frac{\Delta }{2}+\frac{1}{2} i (\omega -l),\frac{\Delta }{2}+\frac{1}{2} i
   (\omega+l );i \omega +1;1-\frac{1}{u^2}\right)\,,
\end{equation}
\begin{equation}
{\cal C}_{\omega l\Delta}\equiv \frac{\Gamma \left(\frac{\Delta }{2}+\frac{1}{2} i (\omega -l)\right) \Gamma \left(\frac{\Delta }{2}+\frac{1}{2} i (\omega +l)\right)}{\Gamma (\Delta -1) \Gamma (i \omega +1)}\,,
\end{equation}
with ${\cal C}_{\omega l\Delta}$ defined for future convenience, $_2F_1$ the Gauss hypergeometric function and ${\cal N}_{\omega l}$ fixed by imposing orthonormality of the KG product\footnote{ \label{Foot7} The orthonormalization of fields on a foliation ending at a horizon is subtle. This is however not related to our concrete problem and has already been extensively covered in the literature, see for example \cite{Kenmoku}.}
\begin{equation}\label{orthonormality}
    (\phi_{\omega l},\phi_{\omega' l'}) = \delta(\omega-\omega') \delta_{l l'},
\end{equation}
so that
\begin{equation}\label{Lmodes}
\left(\square -m^2\right)\phi_{\omega l}(u,\chi,t)=0\; \qquad \partial_t \phi_{\omega l}(u,\chi,t) = - i \omega\; \phi_{\omega l}(u,\chi,t) \qquad \omega>0.
\end{equation}

In $2+1$ dimensions, both Rindler-AdS and BTZ spacetimes are examples of equipartite gravitational systems and their metrics share the form \eqref{metric-example}, albeit the $\chi$ coordinate covering $\mathbb R$ and $S^1$ respectively. Thus, despite being physically distinct, one can effectively follow from this example the analogous BTZ construcion by replacing $\int dk \to r_S^{-1}\sum_{k\in\mathbb{Z}}$ above, where $r_S$ is the horizon radius and $r_S^{-1}$ is the precise factor that maintains the orthonormalization. 

With these eigenfunctions we can calculate precisely the bulk modular flow \eqref{bulkmod} by giving some specific $\lambda(\chi,\tau)$.
In order to show an example we choose a delta-like excitation exactly at $\pi/2$, i.e. $\lambda(\chi,\tau)=\epsilon\,\delta(\tau-\pi/2)e^{i l_0 \chi}$, where  $\epsilon\ll1$ is an dimensionless small parameter that controls the excitation, which leads to
\begin{align}\label{lambda-ej}
    \lambda_{\omega,l} 
    & \equiv \lim_{u\to\infty} u^{\Delta} \int dx \int_0^\pi \lambda(\chi,\tau) \phi_{\omega l}(u,\chi,-i\tau) = \epsilon\, \delta_{l,l_0} {\cal N}\!\!_{\omega l_0} e^{-\omega \pi/2} \left(\alpha_{\omega,l_0,\Delta}\beta_{\omega,l_0,\Delta}-\alpha_{-\omega,l_0,\Delta}\beta_{-\omega,l_0,\Delta}\right),
\end{align}
and then
\begin{align}
    f_\lambda(u,\chi,s)&=\sum_l \int d\omega \lambda^*_{\omega,l} \phi_{\omega l}(u,\chi,s) \\
    &= \epsilon \int d\omega |{\cal N}\!\!_{\omega l_0}|^2 \left(\alpha_{\omega,l_0,\Delta}\beta_{\omega,l_0,\Delta}-\alpha_{-\omega,l_0,\Delta}\beta_{-\omega,l_0,\Delta}\right)^* e^{-\omega \pi/2-i\omega s+i l_0 \varphi} \left[f_{\omega l_0}(u)-f_{-\omega l_0}(u)\right]\;,
\end{align}
in terms of which the bulk modular flow \eqref{bulkmod} can be computed, as well as boundary modular flows via \eqref{boundarymod}. For this particular example we have that
\begin{equation}
\lim_{u\to \infty }\,|u|^\Delta\; \left[ f_\lambda (u,\chi,s) - f_\lambda (u,\chi,0) \right]=  \epsilon \int d\omega |{\cal N}\!\!_{\omega l_0}|^2 |\alpha_{\omega,l_0,\Delta}\beta_{\omega,l_0,\Delta}-\alpha_{-\omega,l_0,\Delta}\beta_{-\omega,l_0,\Delta}|^2 e^{-\omega \pi/2+i l_0 \varphi}[e^{-i\omega s}-1], 
\end{equation}
alongside the generic ($\lambda$-independent) operator piece
\begin{align}
\lim_{u\to \infty }\,|u|^\Delta\; \Phi(u,\chi,s)&= \lim_{u\to \infty }\,|u|^\Delta\; \sum_{\omega>0 l} a_{\omega l} \phi_{\omega l}(u,\chi,s) + h.c. \\
&= \sum_{\omega>0 l}  {\cal N}\!\!_{\omega l} e^{-i\omega s+i l \chi } \left(\alpha_{\omega,l,\Delta}\beta_{\omega,l,\Delta}-\alpha_{-\omega,l,\Delta}\beta_{-\omega,l,\Delta}\right) a_{\omega l} + h.c.
\end{align}

Although $\Phi$ and $\Tilde{\Phi}$ are independent operators in commuting algebras, their respective action on vacuum state are related by an imaginary time translation though the Euclidean piece $E^-$ (see Figs \ref{Fig:Camino}(b) and \ref{Fig:BSMs}(a)), i.e.

\begin{equation}\label{TFDconstraint}
 {\tilde \Phi}(-|u|, t=T_-, \chi)|\Psi_0\rangle = \Phi(|u|, t=T_- -i\pi, \chi)|\Psi_0\rangle,\;~~~~~~ \forall u, \chi\; 
\end{equation}
which must be complemented with a similar condition for the canonically conjugated momentum fields $\Pi(u, t, \chi)$, and so for any operator $A(\Phi, \Pi)$ of the theory.
These equations constitute a constraint to be imposed on the (initial) state at the spacelike surface $t=T_-$.
Recall that the (imaginary) time translation is realized by the operator  $U_0(-i\pi)$, which in the Rindler space is the boost generator \cite{Bisognano}, analytically extended to a purely imaginary parameter.

We have shown in Sec. \ref{modularqft} that this constraint (on the vacuum) is equivalent to the Tomita-Takesaki formalism. In this example we want to see how this also determines the Bogoliubov transformation relating the particle notion for inertial/accelerated observers, and also captures the so-called Unruh trick. In fact, using the (2nd quantized) solution \eqref{quantized-field} and the orthonormality relations of the eigenfunctions $\phi_{\omega l}(u, t, \chi)$, one obtains the following constraint equations
\begin{align}
\hat d^{(1)}_{\omega l}|\Psi_0\rangle \equiv  C_1 \left( \tilde{a}_{\omega l}-e^{-\omega\pi}  a_{\omega l}^\dagger \right)|\Psi_0\rangle =0   \;\qquad
\hat d^{(2)}_{\omega l}|\Psi_0\rangle\equiv C_2 \left( \tilde{a}_{\omega l}^\dagger-e^{+\omega\pi}  a_{\omega l}\right)|\Psi_0\rangle = 0\;, \forall {\omega l}\;,\label{constraint0}
\end{align}
where $a_{\omega l}$ and $\tilde{a}_{\omega l}$ denote the L and R independent ladder operators in ${\cal A}$ and $\tilde{{\cal A}}$ respectively, and $C_{1,2}$ are numeric factors determined by the relations of orthonormality \eqref{orthonormality}. Since these equations can be viewed as annihilating the global vacuum, this procedure defines the Bogoliubov transformation between the $R/L$ ladder operators and the new set $d^{(1,2)}_{\omega l}$, associated to particles for (inertial) observers that have access to the global spacetime. 
One can easily verify that these equations are satisfied by using the explicit form of the state \eqref{Psi0}\footnote{Different formulations of the thermal state condition as a constraint in the string context can be found in \cite{Boots-StringTFD,BootsBrazil2}}.

Therefore, the eigenfunctions associated to these operators, are the precise linear combinations appearing in \eqref{constraint0} of the original $\phi_{\omega l} , \tilde{\phi}_{\omega l}$ solutions, \emph{are analytic} at the throat $u=0$:
\begin{equation}\label{hmodes}
h^{(1)}_{\omega l} = \frac{1}{\sqrt{2 \sinh(\pi \omega)}} \begin{cases} e^{\pi \omega/2} \; \phi^*_{\omega l} & \text{on L}\\ e^{-\pi \omega/2}\; \phi^*_{\omega l} & \text{on R}\end{cases} 
\qquad\qquad 
h^{(2)}_{\omega l} = \frac{1}{\sqrt{2 \sinh(\pi \omega)}} \begin{cases} e^{-\pi \omega/2} \; \phi_{\omega l} & \text{on L}\\ e^{\pi \omega/2} \; \phi_{\omega l} & \text{on R}\end{cases}
\end{equation}
In other words, the correct global canonical quantization of the fields in the manifold lead directly to the analytic global modes defined via the Unruh trick. All these are equivalent restatements of the constraint \eqref{TFDconstraint}. 

The last important aspect of the present example is to show how this constraint/Tomita-Takesaki theory can be generalized to the excited states studied in the paper. If one perform the time translation in $-i\pi$ of the R fields with the sourced evolution operator $U_\lambda(-i\pi)$ in place of $U_0$,
\begin{equation}
\Phi(T_- -i\pi) \equiv U_\lambda(-i\pi)\,\, \Phi(T_-) \,\,U_\lambda (i\pi)\,\,,
\end{equation} 
the constraint \eqref{constraint0} generalizes to 
\begin{equation}\label{TFDconstraint-lambda}
\left[{\tilde \Phi}(-|u|, t=T_-, \chi) - U_\lambda(-i\pi)\,\, \Phi(T_-) \,\,U_\lambda (i\pi) \right]|\Psi_\lambda \rangle =0,\;~~~~~~ \forall u, \chi.\;
\end{equation}

As shown in Sec. \ref{modularqft}, this equation has the ingredients to construct the Tomita-Takesaki theory for excited states. It decomposes in two linearly independent set of equations:
\begin{align}
 \left( \tilde{a}_{\omega l} - e^{-\omega\pi}  a_{\omega l}^\dagger- e^{-\omega\pi} \lambda_{\omega l} \right)|\Psi_\lambda\rangle=0  ~~;~~~~
\left( \tilde{a}_{\omega l}^\dagger - e^{+\omega\pi}  a_{\omega l} - e^{+\omega\pi}\lambda^*_{\omega l} \right)|\Psi_\lambda\rangle = 0\;\;, \forall \omega ,l\;,\label{constraintlamda}
\end{align}
 where we have used that the operator $U_\lambda$ act on ladder operator as a displacement, composed with time translation:
$U_\lambda(-i\pi) \, a^{}_{\omega l} U_\lambda(i\pi) = e^{+\omega\pi}(a^{}_{\omega l} + \lambda^{}_{\omega l}$) (and its h. c.), where the numbers $\lambda_{\omega l}$ are given by \eqref{lambda-n}. It is straightforward to verify that the solution of this equation is the state  \eqref{Psi1}, that can also be expressed as \eqref{psi-desplazamiento}.

Notice finally that these equations can be written as equations of eigenvalues for the new (global) annihilation operators. Multiplying them  by $C_{1,2}$ respectively, we obtain
\begin{equation}\label{constraint-phi}
\left ( \,\hat{d}^{(1,2)}_{\omega l} - \lambda^{(1,2)}_{\omega l} \,\right)\, |\Psi_{\lambda}\,\rangle=0 ~ ~~,  
\end{equation}
where the eigenvalues are given by $\lambda_{\omega l}^{(1)}= C_1 e^{-\omega \pi} \lambda_{\omega l}^*$ and $ \lambda_{\omega l}^{(2)}= C_2 e^{\omega \pi} \lambda_{\omega l}$.
This is nothing but the condition solved by a coherent state of $d$-particles.

It would be interesting to study this construction in other partitions of the system where the operators involved in the TT theory are known. For instance, one could apply an isometry of this spacetime such that the entanglement wedge be dual to the causal development of a ball shaped region in the boundary \cite{Casini2011}.

\subsection{The gravity dual of Modular Hamiltonians for arbitrary entangling surfaces}
\label{Sec:AdS-CFT}

As we mentioned many times the states studied here are particularly relevant in holography since they are closely related to coherent states in the bulk. 
The AdS/CFT conjecture prescribes that the respective Hilbert spaces are equal; therefore, one actually has a single object $|\Psi\rangle$ in a Hilbert space representing the same state. Of course, this state can have very different representations in one or other theory. This hypothesis has been useful to obtain the explicit descriptions of holographic excitations in both theories and to obtain conclusions on their coherence in the bulk at large $N$.

Consequently, by tracing  carefully to both sides of the correspondence (subtleties with the entanglement wedge and the rule to separate in direct products in the bulk should be taken into account \cite{Dong2018}), one obtains that the reduced density matrices also coincide. Thus one concludes that
\begin{equation} \label{Kequals}
K^{CFT} = K^{bulk}
\end{equation}
holds, even thought that the bulk Modular Hamiltonian has a non trivial structure that comes from an expansion in the Newton constant \cite{Aitor2013}, and the purely gravitational contribution $o\left(G_N^{-1}\right)$ involves an \emph{area} operator \cite{JLMS,botta}.

The objective of this section is to take advantage of this formula and use our previous knowledge on excited states in order to compute the contribution of the deformation \eqref{Upm} at $o(G_N^0)$. In principle, this can be used to compute the leading contributions to the (bulk) matrix elements of  $\rho_\lambda[\Sigma_A]$ for any set $A\equiv \partial \Sigma_A$, although in absence of a bulk killing vector one cannot describe the whole (euclidean) space time as $S^1\times \Sigma_A$ and it is hard to check important symmetry features of the Modular Hamiltonian.

The JLMS prescription for the Modular Hamiltonian in a theory consisting of gravity and a nearly free real field $\phi$ is \cite{JLMS, Hayward19}
\begin{equation}
 K_\lambda^{bulk} = \frac{{\hat A}}{4 G_N} + K_\lambda^{grav} + K_\lambda^{matter} 
\end{equation}
where $\hat A$ is the area operator. 
This formula can be obtained from a saddle point approximation (large $N$) of the path integral \eqref{density-bulk} (see Fig. \ref{fig:evo2}).
The (first) area term can be explained from an additional contribution to the boundary term of the gravitational action called Hayward term\footnote{It is the contribution associated to the blue line in Fig \ref{fig:evo2}.} \cite{Hayward1993}, in particular, it was recently shown that the holographic gravitational entropy can be obtained from this term using replica calculations \cite{Hayward19}. We leave the study of this term in the calculus of the Modular Hamiltonian for another work \cite{Botta-Hayward}.

Interestingly, even thought the whole euclidean spacetime ${\cal E}={\cal E}^+ \cup {\cal E}^-$ cannot be foliated as $ S^1_{(\zeta)}\times \Sigma_A$ as in the previous subsections, the 
 matrix elements of the second and third term can be evaluated as 
\be \label{Kgrav}
\langle + |K_\lambda^{grav}| -\rangle =\frac{1}{8\pi G}\,\int_{\Sigma_+} \, \kappa^+ \sqrt{h^+} + \frac{1}{8\pi G}\,\int_{\Sigma_-} \, \kappa^- \sqrt{h^-}
\ee

\be \label{Kmatter} 
\langle+| K_\lambda^{matter} |-\rangle=\,\int_{\Sigma_+} \, \phi^+\Pi^+ \sqrt{h^+} + \int_{\Sigma_-} \, \phi^-\Pi^- \sqrt{h^-} \,+\,\int_{\partial {\cal E}} \, \lambda \partial_{\hat{n}}\lambda\sqrt{h}\,,
\ee
where $h^{\pm}$ are the induced metrics on $\Sigma_{\pm}$ and $\kappa^{\pm}$ their respective extrinsic curvature. 
For concreteness, these expressions are understood in the set up of section \ref{states},
where $|\pm\rangle\equiv |\phi^\pm, h^{\pm}\rangle$ are arbitrary configurations of the fields and induced metrics on the surfaces $\Sigma^\pm$, that are two homologous copies of $\Sigma_A$, as shown in Fig 1b. The asymptotic source $\lambda$ is a smooth function defined on $E^- = \partial {\cal E}^-$ (vanishing on $\tau =0$ and $\tau= -\infty$ for technical issues) and extended to $\partial{\cal E}^+$  with reflection symmetry with respect to $\tau=0$, and $\hat{n}$ is the normal vector to the asymptotic boundary. The solution for the field is

\be\label{solution-packman}\Phi (x) =   \int_{\Sigma^\pm}  G_\pm(x - y) \, \phi^\pm(y) \, dy\,+\,  \int_{\partial {\cal E}} G_\partial(x - z) \, \lambda(z)\, dz
\ee
where $x$ is any point in the bulk and $z\equiv  (\tau , \Omega) \in (-\infty, \infty) \times S^d = \partial {\cal E}$ and $y\in\Sigma^\pm$ . Here $G_\pm$ and $G_\partial$ differ from the standard bulk-to-bulk and bulk-to-boundary propagators. They are solutions to be determined by demanding the following consistency (boundary) conditions.

Denote by $\hat{{\cal E}}$ the euclidean manifold of Fig. \ref{fig:evo2}(b)., then $B_{i}\;,\, i= \Sigma_-, \Sigma_+, \partial {\cal E}$ denotes the three different components of $\partial \hat{{\cal E}}$, and the solution can be expressed as
\be\Phi (x) =  \sum_i \int_{B_i}  G_{i}(x - y) \, \phi_i(y) \, dy\,
\ee
where $\phi_{\partial {\cal E}}(z) \equiv \lambda(z) $, thus, the consistency condition adopts the simple form of boundary conditions 
\begin{equation} \label{PropCond}
G_i(x - y) = \delta_{ij} \;\delta(x - y)    \qquad \text{where } \;x\in B_j\;, \; y\in B_i\;\quad \forall i, j\;.
\end{equation}

Finally, inserting $\Pi^\pm (x) \equiv \pm \partial_\tau \Phi(x)\, |_{\Sigma_\pm}$ and $\Phi^\pm(x) \equiv \Phi(x)|_{\Sigma_\pm}$ into eq. \eqref{Kmatter} we obtain the explicit matrix element in the large $N$ approximation. Observe that this is a quadratic form in the input functions $\Phi^\pm (x)$ , $\Pi^\pm (x)$ and $\lambda(x)$. For a non backreacting field the formula \eqref{Kgrav} can be explicitly calculated in the same way, from the aAdS solution of the Einstein equations on the manifold $\hat{{\cal E}}$  with boundary conditions $h^\pm$ on $\Sigma_\pm$, and  then $\kappa^\pm $ are the (trace of) extrinsic curvatures on these surfaces. 

Let us compute this in the (Euclidean) $2+1$ dimensional AdS spacetime
\be\label{metric-example2}
ds^2=+u^2d\tau^2+\frac{du^2}{1+ u^2}+\left(1+u^2\right)\,d\chi^2 \qquad  \tau\in (-\pi, \pi)\;,\; u\geq 0 \;,\; \chi\in\mathbb{R}\,.
\ee
In order to study the matrix elements we will fix the surfaces $\Sigma^\pm$ on  $\tau=\pm\pi$ to impose the Dirichlet BCs $\phi^\pm$.
In this example we will be able to obtain the required propagators as well as an explicit computation of Modular Hamiltonian matrix elements 
The required propagators are 
\begin{align}\label{NN2}
\int_{\partial {\cal E}} G_\partial( u, \tau,\chi;\tau',\chi') \, & \lambda(\tau',\chi')\, d\tau'dx' =\nn \\&=\int \left( \frac{1}{4\pi} \sum_{l\in\mathbb{Z}} \sum_{m\in\mathbb{Z}}   \;\sin(m \tau )\sin(m \tau' ) e^{i l (\chi-\chi') }  f(-i m,l,u)\right) \lambda(\tau',\chi') d\tau' d\chi',
\end{align}
\begin{align}\label{N1}
 \int_{\Sigma^\pm}  G_\pm(u,\tau,\chi;u',\chi') \, &\phi^\pm(u',\chi') \, du'd\chi' =\nn\\&=\int \left( \frac{i}{4\pi }\sum_{l\in\mathbb{Z}} \sum_{m\in\mathbb{Z}}   \;\sin\left((m+\frac 14)(\tau\pm\pi)\right)
e^{i l (\chi-\chi') }  \phi_{ml}(u) \phi_{ml}(u') \right) \phi^{\pm}(u',\chi') du' d\chi',
\end{align}
where both $f$ and $\phi_{ml} \equiv \phi_{\omega=m,l}$ are defined in \eqref{f} and \eqref{g}.
Note that neither propagator are the standard ones because by frequency quantization they are forced to meet $G_\partial(\tau=\pm\pi)=0$ and $G_{\pm}(u\to\infty)=G_{\pm}(\tau=\mp\pi)=0$ in agreement with condition \eqref{PropCond}. One can explicitly use \eqref{NN2} and \eqref{N1} to compute \eqref{Kmatter}. For the sake of simplicity we pick single mode sources
\begin{equation*}
    \phi^{\pm}(u',\chi')=\phi_{m_{\pm}l_{\pm}}(u)e^{i l_{\pm} \chi'} \;\qquad\qquad \lambda(\tau',x')= \epsilon\; \sin(m_\partial\tau') e^{il_\partial \chi'}\;
\end{equation*}
where again $\epsilon\ll1$ controls the excitation, a straightforward computation leads to
\begin{align}\label{matrixelement}
    \langle m_+, l_+| K_{\lambda_{m_\partial,l_\partial}}^{matter} |m_-, l_-\rangle&= \epsilon\; \delta_{l_\partial,l_+} (-1)^{m_\partial} m_\partial \left[\int du\;u^{-1}\; \phi_{m_{+}l_{+}}(u) f(-im_\partial,l_\partial,u)\right]  + \delta_{l_-,l_+}\delta_{m_-,m_+} (m_-+1/4)\nn\\
    &\quad+\epsilon\; \delta_{l_\partial,l_-} (-1)^{m_\partial} m_\partial \left[\int du\;u^{-1}\; \phi_{m_{-}l_{-}}(u) f(-im_\partial,l_\partial,u)\right]  + \delta_{l_-,l_+}\delta_{m_-,m_+} (m_++1/4)\nn\\
    &\quad+ \epsilon\; {\cal P}_{m_\partial,l_\partial} + \epsilon\; \delta_{l_\partial,l_\pm} \frac{m_\partial (-1)^{m_\partial}}{\left(\frac{1}{4}+m_\pm\right)^2-m_\partial^2} {\cal C}_{nl} \Delta \left(\alpha_{m_\partial,l,\Delta}\beta_{m_\partial,l,\Delta}-\alpha_{-m_\partial,l,\Delta}\beta_{-m_\partial,l,\Delta}\right),
\end{align}
where 
\begin{equation}
    {\cal P}_{m_\partial,l_\partial}=\frac{2(\Delta-1)}{4\pi i }\left(\frac{-1}{e^{2\pi m_\partial}-1}\alpha_{m_\partial,l,\Delta}\beta_{m_\partial,l,\Delta}+\frac{e^{2\pi m_\partial}}{e^{2\pi m_\partial}-1}\alpha_{-m_\partial,l,\Delta}\beta_{-m_\partial,l,\Delta}\right),
\end{equation}
and
\begin{equation}\label{alpha}
\alpha_{\omega l\Delta}\equiv (-1)^{\Delta -1} \frac{ \left(\frac{2-\Delta }{2}+\frac{i}{2}  (\omega - l)\right)_{\Delta -1} \left(\frac{2-\Delta }{2}+\frac{i}{2}  (\omega + l)\right)_{\Delta -1}}{(\Delta -2)! (\Delta -1)!}\,,
\end{equation}
\begin{equation}\label{beta}
\beta_{\omega l\Delta}\equiv-\psi \left(\frac{\Delta }{2}+\frac{i}{2}  (\omega -l)\right)-\psi \left(\frac{\Delta }{2}+\frac{i}{2}  (\omega +l)\right)\,.
\end{equation}
Here $(x)_y$ and $\psi(x)$ are the Pochhammer symbol and the Digamma function respectively. We find the first, third and last term in \eqref{matrixelement} the most relevants because they explicitly show the excited nature of the state. Note also that these are non diagonal pieces of the operator. One can show that the $u$ integrals in brackets are convergent both at $u=0$ and $u\to\infty$, albeit $u=0$ requires careful regularization \cite{Kenmoku}, see footnote \ref{Foot7}.  Considering linearized Einstein gravity, the computation of \eqref{Kgrav} is similar since it involves the same structure and propagators that for the matter field.

We can think the equations \eqref{Kgrav} and \eqref{Kmatter} as providing the natural \emph{candidate} to the gravity dual of Modular Hamiltonians (up to $o\left(G_N^0\right)$) for arbitrary regions $A$ and states $|\Psi_\lambda\rangle$. Nevertheless, in absence of a Killing vector associated to the modular evolution in the bulk, it is difficult to meet the Tomita-Takesaki structure. In the example studied here the $\tau$-dependence of the source $\lambda(\tau)$ manifestly breaks this $U(1)$ symmetry, however a promising method was suggested for these cases in \cite{Botta-Hayward} by considering the calculus for $n$-replicas, and then the modular flow is determined by the analytical extension of $n$ to purely imaginary values.


\section{Summary and conclusions} \label{conclu}

In this paper we studied the Modular Hamiltonian and Modular Flow of a family of excited states whose holographic description is precise in both sides of the AdS/CFT duality and are related to bulk coherent states at large $N$ \cite{Peter 1, Peter 4}. This analysis also captures the complete vacuum sector of the Hilbert space, which are holographically associated to global $n$-particle excitations. 
 
These generating (holographic) states can be constructed geometrically by analytically extending the spacetime to Euclidean times in a Hartle-Hawking fashion, and sourcing the theory with operators on these regions. 
In this set-up, we are able to find Modular Hamiltonian candidates for these systems using a path integral approach. By using TFD  and Schwinger-Keldysh techniques, we manage to frame our excited system as a Tomita-Takesaki theory, allowing us to find the correct $\Delta$ and $J$ operator of our excited system, matching the expressions derived via path integral methods. 
We have shown that when one considers the (extended) modular flow $\Delta^{is}$, a nice geometric structure combining both spacetime signatures emerges, and the Tomita-Takesaki theory can be interpreted geometrically. 
In the case of CFTs, our results can be extended to other bipartite systems related to ours via a conformal map, e.g. a spherical entangling region can be described via the so called CHM map \cite{Casini2011}, see App \ref{App:Ball}. 

It is remarkable that the connection between the TT theory and the TFD formalism, where the so-called thermal state condition is a constraint defining the (thermal) vacuum, 
can be generalized to the holographic states.  
 The vacuum constraint plays an important role in formulating the Unruh problem correctly, and to find the correct Bogoliubov transformation between local and global DOFs. The excited constraint then characterizes simultaneously the state and the action of the operators on it. In terms of the TT theory, the excited constraint can be seen as a deformation of both the vacuum state and modular operator such that the constraint still holds. 
This suggests an interesting
way of interpreting the TT formalism as a constraint between operators of an algebra ${\cal A}$ and $\widetilde{\cal A}$, as they act on a specific state.  
  
By using holography, we are able to study bulk Modular Hamiltonians and their Modular Flows while also retaining the Tomita-Takesaki structure at large $N$. The Modular Hamiltonian for the excited states consists of certain canonical transformation of the original fields and momenta.
The result \eqref{Hlambda} is in agreement
with the result (4.20) in \cite{Liu}, achieved by using AQFT techniques. We present a AdS$_{2+1}$/CFT$_{1+1}$ example in which the explicit modular flow can be computed and within the same example we develop on the relation of the TT theory, TFD formalism and the so called Unruh trick, in order to provide deeper physical insight for the excited state constraint. It is worth to emphasize that this method implicitly assumes the dual map between the objects (operators) of the TT theorem; consequently, the TT construction in aAdS spacetimes implies the one in the strongly coupled CFT.

We also found a formula for the holographic dual of the Modular Hamiltonian for arbitrary spacelike regions $A$ and for an arbitrary coherent excitation $\lambda$. 
Interestingly, the prescription does not rely on the existence of a timelike Killing symmetry associated to the geometric flow
and this would be the natural candidate for the Modular Hamiltonian in the bulk at large $N$. The final expression is non-local, involving special bulk propagators, quadratic in the field on the bulk entanglement region $\Sigma_A$ and in the parameter $\lambda$, which resembles some previous results for free QFTs on a Minkowski spacetime, see \cite{RauloLocal} and references therein. In this case, we also study the example on a bipartite AdS$_{2+1}$ system,
where these special bulk propagators can be explicitly obtained and the matrix elements  \eqref{Kgrav} and \eqref{Kmatter} can be computed.

Finally, the statement \eqref{Kequals} allows to argue (using the BDHM prescription \cite{BDHM}) that the formula \eqref{conjecture}
might be considered a holographic prescription to compute modular evolution of operators in a field theory. It would be interesting to check if the results of Sec.\ref{modbulk}, agrees with an explicit computation with the modular flow in the field theory.

\section*{Acknowledgments}

The authors want to thank Horacio Casini, Diego Pontello, Guillermo Silva, Gonzalo Torroba and Mark Van Raamsdonk for fruitful discussions. Part of the present paper was made on ICTP during the 2019 Spring School on Superstring theory and related topics. Work supported by UNLP and CONICET grants X791, PIP 2017-1109 and PUE B\'usqueda de nueva F\'\i sica.

\begin{appendices}

\section{TFD Basics}{\label{tfd}}

The Thermo-Field Dynamics (TFD) formalism was originally built to study finite temperature QFT in real time using zero temperature techniques \cite{Takashi}. 
In this appendix we present the relevant aspects of the TFD formalism for this work. 

Consider a quantum field theory, whose states belong to the Hilbert space ${\cal H}$. 
In the TFD formalism, one builds a second copy of the system, namely $\widetilde{{\cal H}}$, so that the total system lives in the direct product of the original CFT times its TFD copy, ${\cal H}\otimes\widetilde{{\cal H}}$. 
Thus, given an operator $A$, acting on $\cal H$, one \emph{builds} \cite{Israel} the corresponding operator $\widetilde{A}$  on $\widetilde{{\cal H}}$ using the so-called ``tilde'' conjugation map \cite{UmezawaTFD,Thermal-Coherent},
\begin{align}\label{Tilde-rules}
[A,\tilde B]=0 && (AB)\tilde{\,}=\tilde A \tilde B &&
(c_1 A + c_2 B)\tilde{\,} = c_1^* \tilde A + c_2^* \tilde B && (A^{\dagger})\tilde{\,}=\tilde A ^{\dagger}\;.
\end{align}
Alternatively, one can denote the extended operators as $A_L$ and $A_R$ respectively:
\be \mathbb{I}\otimes A \equiv A \equiv A_R  \;,\qquad\qquad \tilde (\mathbb{I}\otimes A) = \tilde{A} \otimes \mathbb{I} \equiv \tilde A \equiv A_L\;.
\ee
We will often use both notations alternatively throughout this work. 

One can now define the TFD vacuum, denoted $|\Psi_0\rangle\in{\cal H}\otimes\widetilde{{\cal H}}$ as follows. We start from the identity state
$$|1\rangle\!\rangle \equiv \sum_n |n\rangle\otimes|n\rangle = e^{a_L^\dagger \, a_R ^\dagger}|0\rangle\otimes|0\rangle$$ 
which is an auxiliary maximally entangled state of the energy eigenfunctions of the spaces ${\cal H}$ and $\widetilde{{\cal H}}$ with divergent norm. The (unnormalized) TFD vacuum can be built as,
\begin{equation}
    |\Psi_0\rangle \equiv \sum_n e^{-\frac \beta2 E_n} |n\rangle\otimes|n\rangle = e^{-\frac \beta2 H}|1\rangle\!\rangle=e^{-\frac \beta2 \tilde H}|1\rangle\!\rangle\;.
\end{equation}
where $\beta^{-1}=T$ is the temperature of the system and $H$, $\tilde H$ and $E_n$ are the system and copy Hamiltonians and its energy eigenvalues respectively. Notice that $|\Psi_0\rangle$ is also a maximally entangled state.
The relevance of the TFD vacuum resides in that it allows to compute expectation values at finite temperature of the original system ${\cal H}$ as VEVs in the doubled space ${\cal H}\otimes\widetilde{{\cal H}}$. It can explicitly be checked that \cite{Takashi}, 
\be \label{id2} 
\langle A\rangle_\beta \equiv \text{tr} \;\{\rho A\} =\langle \Psi_0|A\otimes\mathbb{I}|\Psi_0\rangle\;; \qquad\qquad \rho = e^{-\beta H}\;.
\ee
The \emph{vacuum} character of $|\Psi_0\rangle$ can be understood in terms of the global Hamiltonian $(H_R-H_L)\in{\cal H}\otimes\widetilde{{\cal H}}$, for which it is immediate to check
$$(H_R-H_L)|\Psi_0\rangle=0\;.$$
Notice that the systems are decoupled and its interaction is entirely due to the maximally entangled character of the theory vacuum $|\Psi_0\rangle$. 
The equation above suggests a physical interpretation in terms of two systems evolving in opposite time directions. This interpretation has found holographic support especially in the eternal BH solutions \cite{eternal}. It has also been observed that the DOFs splitting of a system into two Rindler patches can be understood as a TFD doubled space \cite{RindlerAdSCFT}. 

\begin{figure}[t]\centering
\begin{subfigure}{0.49\textwidth}\centering
\includegraphics[width=.9\linewidth] {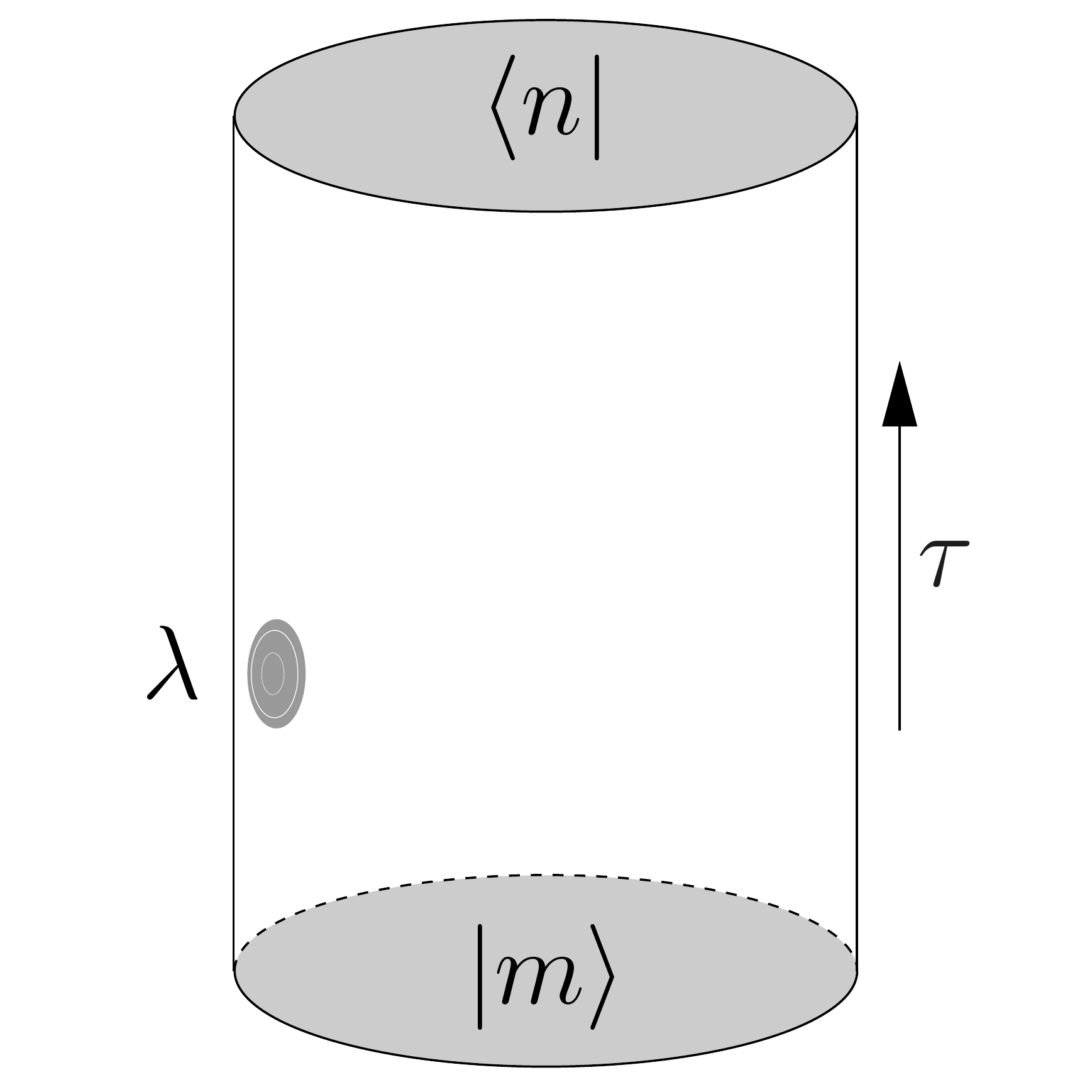}
\caption{}
\end{subfigure}
\begin{subfigure}{0.49\textwidth}\centering
\includegraphics[width=.9\linewidth] {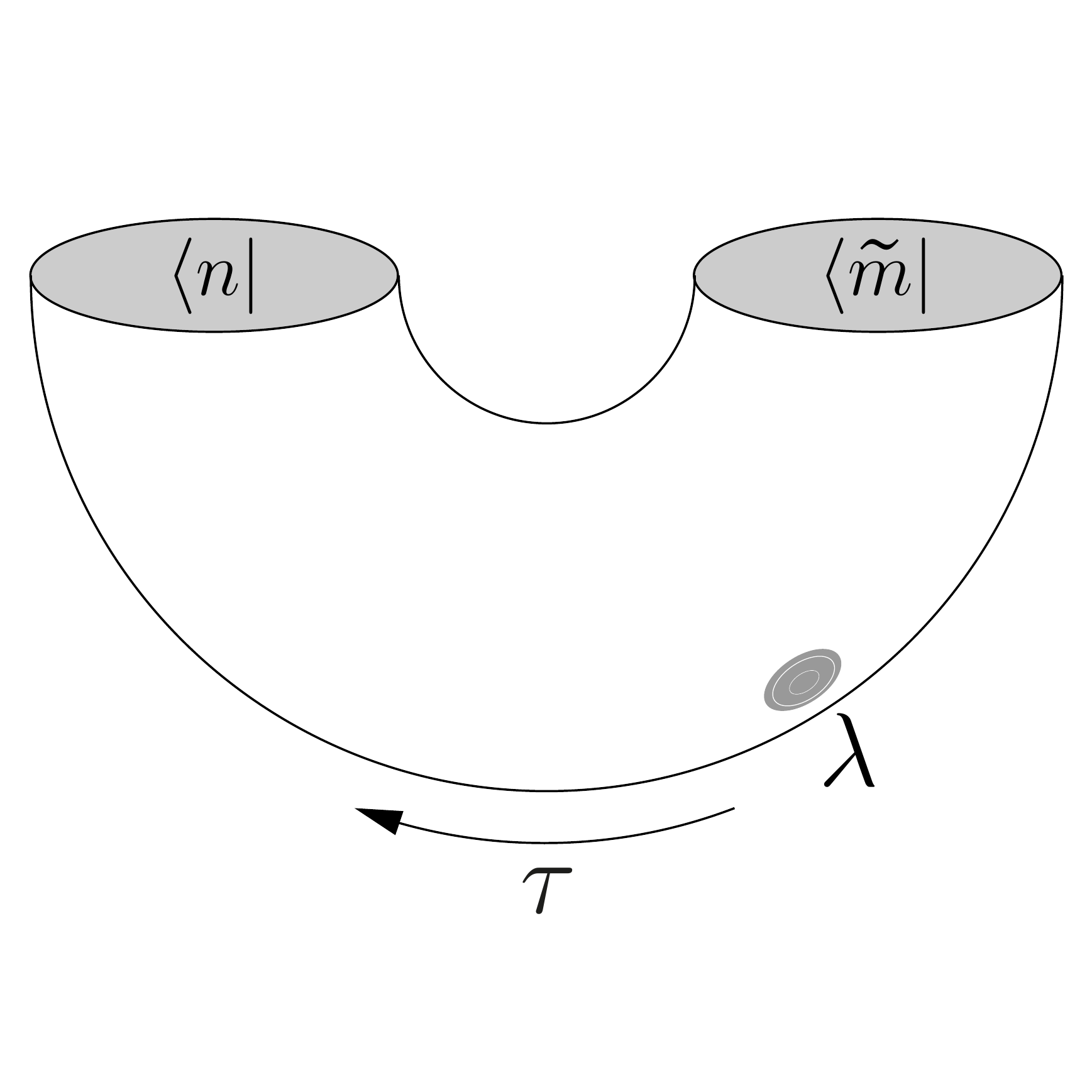}
\caption{}
\end{subfigure}
\caption{(a) A piece of Euclidean evolution cut at regions $\Sigma_1$ and $\Sigma_2$ understood depicted as the matrix element $\langle n|\rho_\lambda| m \rangle$ of a density matrix $\rho_\lambda$. (b) The same geometry can be instead understood as the coefficient $\langle\!\langle n \tilde m |\Psi_\lambda \rangle\!\rangle$ of a ket $|\Psi_\lambda \rangle\!\rangle$ defined in the TFD Hilbert space $\cal H \bigotimes\tilde H$.}
\label{Fig:States}
\end{figure}

In this work, we exploit the fact that the TFD vacuum can also be thought as an Euclidean time evolution operator $U_0(0,i\beta/2)$ acting on the identity state,
\begin{equation}\nn
    |\Psi_0\rangle = U_0(0,-i\beta/2)\otimes\mathbb{I}|1\rangle\!\rangle = \mathbb{I} \otimes U_0(0,-i\beta/2)  |1\rangle\!\rangle\;,
\end{equation}
and study excitations of the TFD vacuum defined as in \eqref{state-U},
\begin{equation}\nn
    |\Psi_\lambda\rangle= U_\lambda(0,-i\beta/2)\otimes\mathbb{I}|1\rangle\!\rangle = \mathbb{I} \otimes U_\lambda(0,-i\beta/2)  |1\rangle\!\rangle\;.
\end{equation}
This equation, projected into an energy eigenstate basis, can be also geometrically understood as shown in figure \ref{Fig:States}: $U_\lambda$ is depicted on the left as an evolution operator on a single Hilbert space, the corresponding TFD-ket  $|\Psi_\lambda\rangle$ is illustrated on the right with the two cylinder's ends now representing the doubled TFD DOFs at some spacelike surface at a fixed time $t$. It is important to notice that the excitation under study is created with an operator that can be fully localized in only one of the factors of the Hilbert space. Finally, a density matrix associated with the state $|\Psi_\lambda\rangle$ can also be defined as
$$\rho_\lambda=U_\lambda (i\pi, 0) U_\lambda^\dagger (i\pi, 0)\equiv U_\lambda (i\pi, 0) U_\lambda (0, -i\pi)=\text{Tr}\{|\Psi_\lambda\rangle\langle\Psi_\lambda|\}\;.$$ 
                                                            
The relationship between the Tomita-Takesaki structure and the TFD construction is well known in the literature, see for example \cite{TFDyTT}. We would like to conclude this section by showing that the states $|\Psi_\lambda\rangle$ are cyclic and separating. It can be shown that $|\Psi_\lambda \rangle$ is cyclic and separating if there are no non-trivial operators in either $\cal H$ or $\widetilde {\cal{ H}}$ such that $B|\Psi_\lambda \rangle = 0$, see \cite{Witten2018}. These are both necessary hypothesis for the Tomita-Takesaki theorem to hold which we use throughout this work.  A more formal introduction to these properties can be found in \cite{Haag}.

Assume that there exists an operator $ \tilde B\in \widetilde{{\cal H}}$ such that
\be\label{B0}
\tilde B|\Psi_\lambda \rangle = (\tilde B  \otimes U_\lambda ) |1\rangle\!\rangle  =0  ,
\ee 
multiply this by $U_\lambda^\dagger$ to get 
\be\label{B}
U_\lambda^\dagger (\tilde B \otimes U_\lambda) |1\rangle\!\rangle = (\tilde B  \otimes \rho_\lambda) |1\rangle\!\rangle  =0 \;.
\ee
Since $\rho_\lambda$ is Hermitian (and positive), it is invertible and can removed from this equation, i.e.
\be
 B |1\rangle\!\rangle  = 0  \,.
\ee
Recalling that $|1\rangle\!\rangle=\sum_n |n\rangle |\tilde{n}\rangle$ has been defined in terms of a complete orthonormal basis of ${\cal H}\otimes\widetilde{{\cal H}}$, we project this equation on an arbitrary element $\langle m |\langle \tilde{k} |$ and obtain $\langle \tilde{k} | B |\tilde{n}\rangle =0$ for all $n, k$, i.e all the matrix elements of the operator $B$ vanish. This shows that $|\Psi_\lambda \rangle$ is cyclic.

Proving that the state is separable, i.e. that the state is cyclic with respect to an operator $B\in{\cal H}$, follows analogously by recalling that the state can also be equivalently written in terms of an operator $\tilde{U}_\lambda \in  \widetilde{{\cal H}}$. This concludes the demonstration.

\section{Excited states in a Ball }\label{App:Ball}

We devote this appendix to show in a concrete example how our results for equipartite systems extend to other bipartite systems related via a conformal transformation. In particular, using the CHM map \cite{Casini2011} we will obtain the excited Modular Hamiltonian for a ball shaped region. We begin by briefly reviewing the sSK construction discussed in section \ref{Sec:sSK}. 
We then follow the CHM map to describe the modular flow in the complexified sSK geometry and conclude by obtaining the excited Modular Hamiltonian of the system.

In the case of the Rindler spacetime,the sSK extension is built 
from the Rindler wedge as follows. One takes the standard Minkowski spacetime that, covered by Rindler coordinates, splits in four regions or patches. Then, let us take only the left and right sides $W\equiv W_L \cup W_R$ whose boundaries $\Sigma_\pm $ are homologous to an extended foliation  $\Sigma(t)$ of $W$ that corresponds to the parameter $t\to\pm\infty $ \footnote{In this and other cases, the associated algebra of operators to $W_L$ is the commuting of algebra of $W_R$.}. This is the real time extended Rindler wedge in the complexified geometry, but it is convenient to consider $W(T_-, T_+) \subset W$  between finite limits of the real time parameter. 
Take two halves of the analytical extension of the Rindler spacetime to purely imaginary time coordinate $t\to -i \tau$. 
The rank of the coordinate $\tau$ must be $[0,2\pi ]$ in order to avoid the conical singularity at the origin.
Now we split this geometry in two (past and future) halves $E^\pm$ by the intervals $\tau \in [0, \pi ]$ and $\tau \in [\pi , 2\pi ]$ respectively so we can define the closed complexified (Rindler) space time, denoted by $W_{\cal C}$ by smoothly gluing $E^\pm$ with $W$ through the surfaces $\Sigma^\pm$.
This construction is similar to \cite{Peter 3,Peter 4} and the smoothness conditions implies the continuity of the metric and the extrinsic curvature along the parameter $\tau$. The total geometry can be seen as a fibration $W_{{\cal C}} = \Sigma_R \times {\cal C}$ and is shown un Fig. \ref{Fig:Camino}.

We now turn to the analysis in the ball.
Let us see, first in a naive way, which should be the explicit form of the Modular Hamiltonian for the excited state on the spherical entangling surface. The strategy will be to do the conformal transformation that maps the Rindler wedge to the causal development of a sphere of radius $R$: $D\equiv D(V)$, this is the so called CHM map \cite{Casini2011}. First, for simplicity, suppose that the source $\lambda$ does not depend on the time coordinate, such as in equation \eqref{KR}. Applying the CHM map (implemented by an operator ${\cal U}$) to both sides of \eqref{KR} we obtain
\be  
{\cal U}\, K_\lambda \, {\cal U}^{-1} =   2\pi\,\left( {\cal U}\, K_0 \, {\cal U}^{-1} + \,\int_\Sigma \lambda( X)  {\cal U}\,  {\cal O}(X)\, {\cal U}^{-1} \;\sqrt{g_\Sigma}dX^{d-1}\,\right)
\ee
Using that $ {\cal U}\,  {\cal O}(X)\, {\cal U}^{-1} = \Omega^{-\Delta}(x) {\cal O}_D(x)$, we obtain the form of the Modular Hamiltonian in the ball (capital $X$ stands for coordinates on the Rindler space and small $x$ for those on the transformed space),
\be\label{guessKD}  K_D =  2\pi\, K_{0_D} + \,\left(\int_{V} \, \lambda(x)  \Omega^{-\Delta}(x) \, {\cal O}_D(x)\,  \beta(x)\, \sqrt{g_V} dx^{d-1}\,\right).
\ee

The symbols $\Delta$ and $\Omega$ here stand for the scaling dimension of the operator and conformal factor introduced by the map respectively, and $K_{0_D}$ was computed explicitly in \cite{Casini2011}. The factor $\beta(x)$ comes from the \emph{dilatation}  of the time coordinate due to the conformal map. This is in agreement with results found in \cite{Sarosi}.

\begin{figure}[t]\centering
\begin{subfigure}{0.49\textwidth}\centering
\includegraphics[width=.95\linewidth] {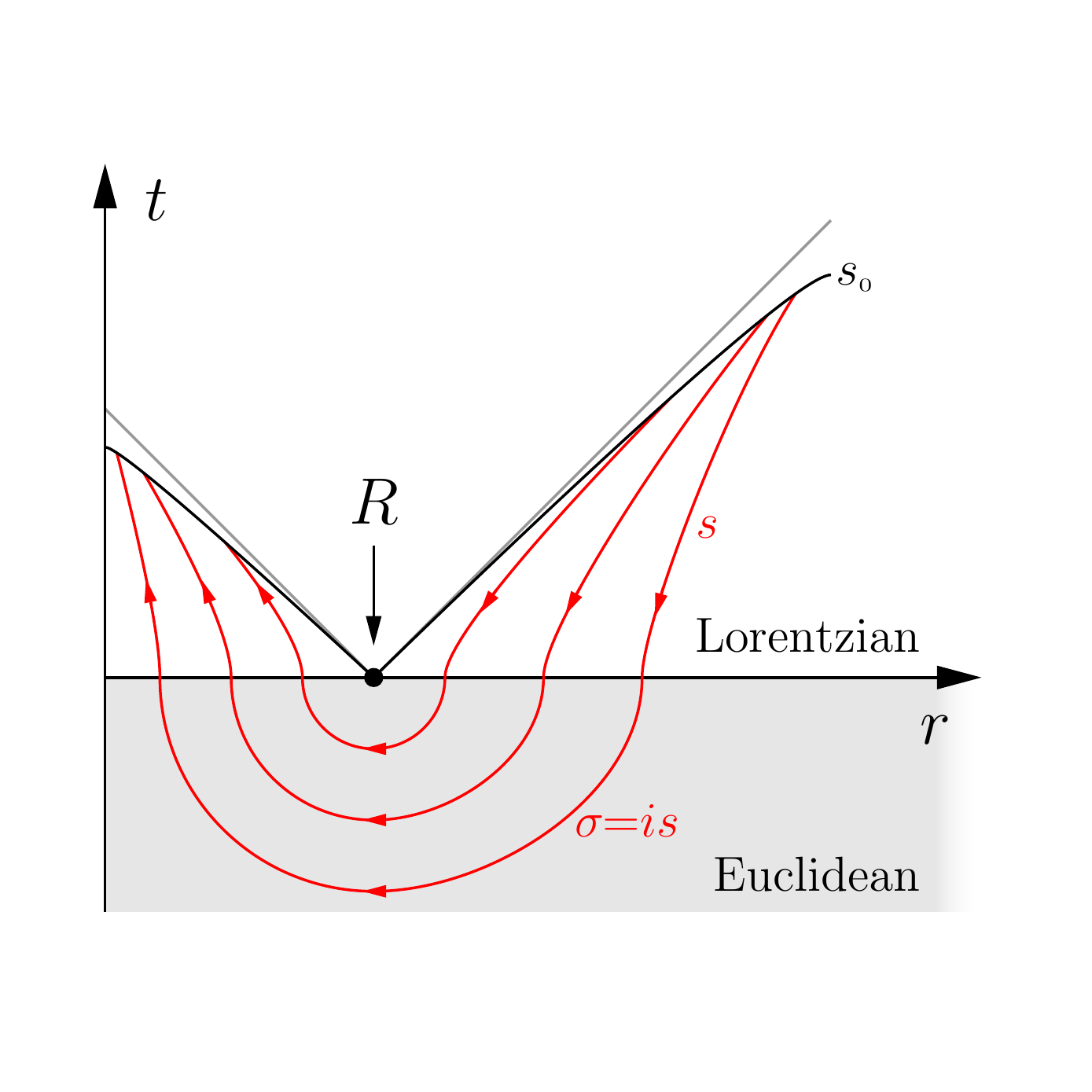}
\caption{}
\end{subfigure}
\begin{subfigure}{0.49\textwidth}\centering
\includegraphics[width=.95\linewidth] {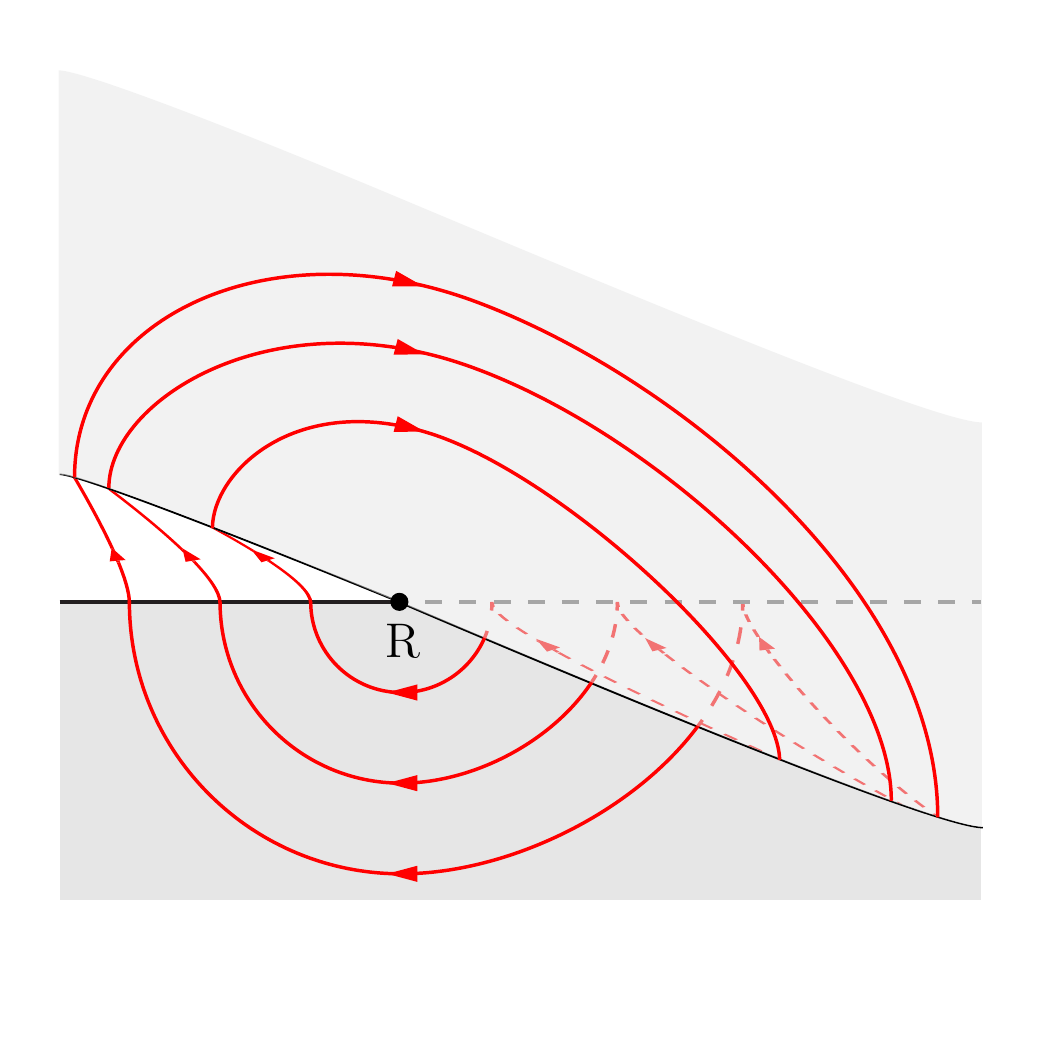}
\caption{}
\end{subfigure}
\caption{ (a) Under the CHM map $\cal U$, one can also build a geometric interpretation of the modular operator and evolution for a ball in a CFT. Notice that the Euclidean evolution resembles a circle near $R$ but departs from this behaviour at greater distances. (b) A closed contour in complex modular evolution can also be geometrically interpreted.}
\label{Fig:BSMs}
\end{figure}

Let us derive this expression from a path integral approach for a general source  $\lambda(x, \tau)$.
The previous construction of the sSK path allows to compute time ordered $n$-point functions in arbitrary points of the extension $W_{\cal C}$,
and then one can also construct the corresponding  sSK extension for the ball, $ D_{\cal C}$, by applying the CHM transformation to each component
of $ W_{l}$, and glue them. Here $l=W_R, E^- , W_L, E^+$ refers to all the pieces of the symmetric SK complexified spacetime.
In particular $W_R, W_L$ map into $D\, , \,D(\bar{V})$ respectively (see figs \ref{Fig:BSMs}). Since the analytical extension of modular flow $x(-is)$ is ill defined for the center of the ball $x^i=0\,\,\, i=1\dots d$, it is convenient to define $D_{\cal C}$ as the foliation $\{V_0(\theta)\}_{\cal C} \sim V_0 \times {\cal C}$ where $V_0$ is the ball minus this point. 

Consider then the sSK extension of the result \eqref{density-rindler-U}
\begin{equation}\label{ZCFT}
Z_{} (\lambda)= \text{Tr} \, \;  U\qquad\qquad  U \equiv {\cal P} \,e^{-i\int_{{\cal C}_{}} d\theta\; (K_0 + {\cal O}. \,\lambda(\theta))},
\end{equation}
then the n-points correlation functions in the Rindler wedge can be computed from
\be\label{n-points-derivadas-rindler}
 \langle \Psi_{0_R}| {\cal O}(X_1){\cal O}(X_2) \dots {\cal O}(X_n)|\Psi_{0_R} \rangle=   (-i)^n \left. \frac{\partial^{n}}{\partial\lambda(X_1)^{}\partial\lambda(X_2)^{}\dots} Z_R(\lambda) \right|_{\lambda = 0} \,
\ee
for all set of (arbitrary) $n$ points $X^\mu_1 , \dots , X^\mu_n\,\in  W_R $.

Now we will apply the CHM map, which is nothing but a conformal transformation $W_R \to D\equiv D(B_R)\,$ implemented by the unitary transformation $\cal{U}$ on the Hilbert spaces. In particular the (scalar) primary operators transform as
\be\label{CHMprimaries}
{\cal{O}}_D(x) = \Omega(X)^\Delta \,{\cal U} \,{\cal{O}} (X) \,{\cal U}^{-1}\,;\qquad\qquad \forall X^\mu \in W_R\;.
\ee
Using that ${\cal U}|0\rangle = |0\rangle $ we obtain
\be\label{n-ponits-pure}
{}_0\langle {\cal O}_D(x_1){\cal O}_D(x_2) \dots {\cal O}_D(x_n) \rangle_0= \prod^n_{i=1}\Omega(X_i)^\Delta {}_0\langle{\cal O}(X_1){\cal O}(X_2) \dots {\cal O}(X_n) \rangle_0
\ee 
for any set of (arbitrary) $n$ points $X^\mu_1(\theta),\dots ,X^\mu_n(\theta)$, at the same hypersurface $\theta =$constant. The left hand side is nothing but 
\be
(-i)^n  \left. \frac{\partial^{n}}{\partial\lambda(x_1)^{}\partial\lambda(x_2)^{}\dots} Z_D(\lambda) \right|_{\lambda = 0},
\ee
so one can think these relations as \emph{probing} the generating function for both theories in both extended spaces. In fact they imply that the expansions (in powers of $\lambda$) of both functionals coincide
\be\label{mapZ}
Z_D(\lambda) = Z_R(\lambda \to\Omega^{-\Delta}\lambda,  W_{\cal C} \to D_{\cal C} ),
\ee
where
\be \label{generating-function-D}
Z_D =\,\int\; [D\phi] \;e^{i \int_{D_{\cal C}} \,d\theta\,\sqrt{-g_V}\,d^d x\left({\cal L}_{CFT} + \Omega^{-\Delta}(x^\mu)\,\lambda(x)\,{\cal O}_D(x)\right)}
\ee 
satisfies all the relations \eqref{n-ponits-pure}. Since $D_{\cal C} = V_0 \times {\cal C}$, this is defined on fields with periodic  conditions in $\theta$, and  we can also express this as 
\begin{equation}\label{ZCFT-Ball}
Z_{D} (\lambda)= \text{Tr} \, \;  U_D\qquad\qquad  U_D \equiv {\cal P} \,e^{-i\int_{{\cal C}_{}} d\theta\; (K_{0_D} + \int_{V_0}\,\sqrt{-g_V}\,d^d x \,\beta(x)\,\Omega^{-\Delta}(x) {\cal O}_D(x) \,\lambda(\theta, x))},
\end{equation}
where the exponent corresponds to the canonical energy for each slice $V^0(\theta)$ computed from the deformed Lagrangian of \eqref{generating-function-D} \footnote{It is the component $T_V^{\mu \nu}  n_\mu(\theta) \tau^{(V)}_\nu (\theta) $ of the energy-momentum tensor derived from  \eqref{generating-function-D}, where $n^\mu(\theta)$ is the unit vector, orthonormal to $V_0 (\theta)$ and recalling that $\theta$ is the analytically extended Rindler time, $\beta(x)$ is locally defined by $\frac{\beta}{2\pi}\,\tau^{(V)}_\nu (\theta) \equiv  \partial_\nu \theta $.}.

The same argument holds for the matrix elements of $U(\lambda)$ (and $U_D(\lambda)$). In fact,
one can remove the  periodic boundary condition from this path integral, and to consider the evolution operator between two hypersurfaces $\Sigma(\theta_1)$ and $\Sigma(\theta_2)$ by
imposing arbitrary field configurations on each one. Then the (dynamical) evolution operators relate by 
\be\label{mapU}
U_D (\theta_1 , \theta_2)[\lambda] = {\cal U}  \, U(\theta_1 , \theta_2)[\lambda] \,{\cal U}^{-1}\;\; 
\ee
and the matrix elements can be computed with the following path integral:
\be 
\langle \phi_1 | U_D(\lambda) |\phi_2\rangle=\int^{\phi_2}_{\phi_1} [D\phi] e^{i \int_{\theta_1}^{\theta_2} \,d\theta\,\int_{V_0 (\theta)}\sqrt{-g_V} \,d^d x\left({\cal L}_{CFT} + \Omega^{-\Delta}(x^\mu)\,\lambda(x)\,{\cal O}_D(x)\right)}\nonumber
\ee
where $\phi_1\,,\, \phi_2$ are two arbitrarily specified configurations of the fields on the surface $V_0(\theta_1), V_0(\theta_2)$ respectively. Since the CHM conformal transformation maps one-to-one the points of $B_R(\theta)$ into $\Sigma_R (\theta)$ (and $\bar{V}(= V_0(-i\pi))$ into $\Sigma_L$) these correspond to the configurations on $\Sigma(\theta_1), \Sigma(\theta_2)$ of the sSK extension of the Rindler wedge. Finally, by taking $\theta_{1, 2}$ to be the red points in Fig 2 (a), and taking trace
we obtain the main formula \eqref{mapZ}. 

By virtue of \eqref{mapU}, we have 
\be\label{mapUl}
U^l_D (\lambda) =  {\cal U}\, U^l (\lambda) \, {\cal U}^{-1}\;\; 
\ee
where $l=W_R, E^- , W_L, E^+$ label the pieces of the symmetric SK complexified spacetime. 
The reduced matrix density of excited global (pure) states can be obtained by taking the limit $|T^+ - T^-| \to 0$ and removing the real time components of the geometry, then the entire sSK geometry is nothing but $E\equiv E^+ \cup E^-$.
In fact, the analytical extension to purely imaginary values of the parameter $t\to -i\tau$, $\tau\in [0,2\pi)$ evolves the operators in the manifold $\Sigma\times S^1$.  The (pure) global state is built with the evolution operator on the interval $(0,\pi)$, so the excited states can be systematically constructed by deforming the CFT action with a source $\lambda(X, \tau)$, therefore by virtue of \eqref{densitystate}, we must extend the source to all the manifold $E$ demanding $\lambda(X, \tau) = \lambda(X, -\tau)$. Because of the CHM map, all these remarks can be transplanted to the description of the CFT on $D_{\cal C}$ with the DOF within a sphere $V$, see fig. \ref{Fig:BSMs}(b).

Using the result \eqref{densitystate}, and using \eqref{mapUl}, we obtain the (unnormalized)  reduced density matrix for any $\lambda$-state in the ball shaped region  
\be
\rho_D(\lambda) = U_D (0, 2\pi) (\lambda)\, ,
\ee 
whose matrix elements are, by virtue of \eqref{mapU},
\be 
\langle \phi_1 | \rho_D(\lambda) |\phi_2\rangle=\int^{\phi_2}_{\phi_1} [D\phi] e^{-\, \int_{0}^{2\pi} \,d\tau\,\int_{V_0 (\tau)}\,\sqrt{-g_V}d^d x\, \left({\cal L}_{CFT} + \Omega^{-\Delta}(x^\mu)\,\lambda(x)\,{\cal O}_D(x)\right)}\nonumber
\ee
A similar formula and construction can be obtained for any region obtained from the Rindler spacetime by some conformal mapping.


\end{appendices}


\begin{thebibliography}{99}

\bibitem{Haag}
  R.~Haag,
  Berlin, Germany: Springer (1992) 356 p. (Texts and monographs in physics)

\bibitem{Witten2018}
  E.~Witten,
  Rev.\ Mod.\ Phys.\  {\bf 90} (2018) no.4,  045003
  doi:10.1103/RevModPhys.90.045003
  [arXiv:1803.04993 [hep-th]].

\bibitem{Takesaki}
M.~Takesaki,
Lecture Notes in Mathematics, Vol. 128. (1970) 

\bibitem{Li}
  H.~Li and F.~Haldane,
  Phys.\ Rev.\ Lett.\  {\bf 101} (2008) no.1,  010504
  doi:10.1103/PhysRevLett.101.010504
  [arXiv:0805.0332 [cond-mat.mes-hall]].

\bibitem{Chang}
  P.~Y.~Chang, J.~S.~You, X.~Wen and S.~Ryu,
  \emph {Entanglement spectrum and entropy in topological non-Hermitian systems and non-unitary conformal field theories,}
  arXiv:1909.01346 [cond-mat.str-el].


  \bibitem{RauloLocal}
  R.~Arias, D.~Blanco, H.~Casini and M.~Huerta,
  Phys.\ Rev.\ D {\bf 95} (2017) no.6,  065005
  doi:10.1103/PhysRevD.95.065005
  [arXiv:1611.08517 [hep-th]];\\ 
  R.~Arias, H.~Casini, M.~Huerta and D.~Pontello,
  Phys.\ Rev.\ D {\bf 96} (2017) no.10,  105019
  doi:10.1103/PhysRevD.96.105019
  [arXiv:1707.05375 [hep-th]];\\
  S.~Hollands,
  doi:10.1007/s11005-019-01238-z
  arXiv:1903.07508 [hep-th];\\
  H.~Casini, S.~Grillo and D.~Pontello,
  Phys.\ Rev.\ D {\bf 99} (2019) no.12,  125020
  doi:10.1103/PhysRevD.99.125020
  [arXiv:1903.00109 [hep-th]];\\
  D.~D.~Blanco, H.~Casini, L.~Y.~Hung and R.~C.~Myers,
  JHEP {\bf 1308} (2013) 060
  doi:10.1007/JHEP08(2013)060
  [arXiv:1305.3182 [hep-th]].

\bibitem{deboer}
  J.~De Boer, J.~Jarvela and E.~Keski-Vakkuri,
  Phys.\ Rev.\ D {\bf 99} (2019) no.6,  066012
  doi:10.1103/PhysRevD.99.066012
  [arXiv:1807.07357 [hep-th]].

\bibitem{Bisognano}
  J.~J.~Bisognano and E.~H.~Wichmann,
  J.\ Math.\ Phys.\  {\bf 16} (1975) 985.
  doi:10.1063/1.522605

\bibitem{Casini2011}
  H.~Casini, M.~Huerta and R.~C.~Myers,
  JHEP {\bf 1105} (2011) 036
  doi:10.1007/JHEP05(2011)036
  [arXiv:1102.0440 [hep-th]].

\bibitem{Cardy}
  J.~Cardy and E.~Tonni,
  J.\ Stat.\ Mech.\  {\bf 1612} (2016) no.12,  123103
  doi:10.1088/1742-5468/2016/12/123103
  [arXiv:1608.01283 [cond-mat.stat-mech]];\\
  X.~Wen, S.~Ryu and A.~W.~W.~Ludwig,
  Phys.\ Rev.\ B {\bf 93} (2016) no.23,  235119
  doi:10.1103/PhysRevB.93.235119
  [arXiv:1604.01085 [cond-mat.str-el]];\\
  X.~Wen, S.~Ryu and A.~W.~W.~Ludwig,
  J.\ Stat.\ Mech.\  {\bf 1811} (2018) no.11,  113103
  doi:10.1088/1742-5468/aae84e
  [arXiv:1807.04440 [cond-mat.str-el]];\\
   G.~Di Giulio, R.~Arias and E.~Tonni,
  J.\ Stat.\ Mech.\  {\bf 1912} (2019) no.12,  123103
  doi:10.1088/1742-5468/ab4e8f
  [arXiv:1905.01144 [cond-mat.stat-mech]].

\bibitem{Arias2018}
  H.~Casini and M.~Huerta,
  Class.\ Quant.\ Grav.\  {\bf 26} (2009) 185005
  doi:10.1088/0264-9381/26/18/185005
  [arXiv:0903.5284 [hep-th]];\\
  R.~E.~Arias, H.~Casini, M.~Huerta and D.~Pontello,
  Phys.\ Rev.\ D {\bf 98} (2018) no.12,  125008
  doi:10.1103/PhysRevD.98.125008
  [arXiv:1809.00026 [hep-th]].
  
  \bibitem{BlancoGuillem}
  D.~Blanco and G.~Perez-Nadal,
  Phys.\ Rev.\ D {\bf 100} (2019) no.2,  025003
  doi:10.1103/PhysRevD.100.025003
  [arXiv:1905.05210 [hep-th]];\\
   P.~Fries and I.~A.~Reyes,
  Phys.\ Rev.\ Lett.\  {\bf 123} (2019) no.21,  211603
  doi:10.1103/PhysRevLett.123.211603
  [arXiv:1905.05768 [hep-th]];\\
  P.~Fries and I.~A.~Reyes,
  Phys.\ Rev.\ D {\bf 100} (2019) no.10,  105015
  doi:10.1103/PhysRevD.100.105015
  [arXiv:1906.02207 [hep-th]].
  

\bibitem{adscft}
  J.~M.~Maldacena,
  Int.\ J.\ Theor.\ Phys.\  {\bf 38}, 1113 (1999)
  [Adv.\ Theor.\ Math.\ Phys.\  {\bf 2}, 231 (1998)],
  [hep-th/9711200].

\bibitem{GKP}
 S.~S.~Gubser, I.~R.~Klebanov and A.~M.~Polyakov,
  Phys.\ Lett.\ B {\bf 428}, 105 (1998), [hep-th/9802109].

\bibitem{W}
E.Witten, Adv. Theor. Math. Phys. 2, 253 (1998), [hep-th/9802150].

\bibitem{Casinibound}
  H.~Casini,
  Class.\ Quant.\ Grav.\  {\bf 25} (2008) 205021
  doi:10.1088/0264-9381/25/20/205021
  [arXiv:0804.2182 [hep-th]].

\bibitem{Faulkner}
  T.~Faulkner, R.~G.~Leigh, O.~Parrikar and H.~Wang,
  JHEP {\bf 1609} (2016) 038
  doi:10.1007/JHEP09(2016)038
  [arXiv:1605.08072 [hep-th]];\\
  S.~Balakrishnan, T.~Faulkner, Z.~U.~Khandker and H.~Wang,
  JHEP {\bf 1909} (2019) 020
  doi:10.1007/JHEP09(2019)020
  [arXiv:1706.09432 [hep-th]].

\bibitem{An}
Y.~An and P.~Cheng,
[arXiv:2004.14059 [hep-th]].


\bibitem{JLMS}
  D.~L.~Jafferis, A.~Lewkowycz, J.~Maldacena and S.~J.~Suh,
  JHEP {\bf 1606} (2016) 004
  doi:10.1007/JHEP06(2016)004
  [arXiv:1512.06431 [hep-th]].
  
  \bibitem{Jafferis}
  D.~L.~Jafferis and S.~J.~Suh,
  JHEP {\bf 1609} (2016) 068
  doi:10.1007/JHEP09(2016)068
  [arXiv:1412.8465 [hep-th]].

\bibitem{Faulkner2}
  T.~Faulkner, M.~Li and H.~Wang,
  JHEP {\bf 1904} (2019) 119
  doi:10.1007/JHEP04(2019)119
  [arXiv:1806.10560 [hep-th]].

\bibitem{Papadodimas2013}
K.~Papadodimas and S.~Raju,
Phys. Rev. D \textbf{89} (2014) no.8, 086010
doi:10.1103/PhysRevD.89.086010
[arXiv:1310.6335 [hep-th]].

\bibitem{Jefferson2018}
R.~Jefferson,
SciPost Phys. \textbf{6} (2019) no.4, 042
doi:10.21468/SciPostPhys.6.4.042
[arXiv:1811.08900 [hep-th]].
\bibitem{Czech2019}
B.~Czech, J.~De Boer, D.~Ge and L.~Lamprou,
JHEP \textbf{11} (2019), 094
doi:10.1007/JHEP11(2019)094
[arXiv:1903.04493 [hep-th]].

\bibitem{deBoer2019}
J.~De Boer and L.~Lamprou,
JHEP \textbf{06} (2020), 024
doi:10.1007/JHEP06(2020)024
[arXiv:1912.02810 [hep-th]].


\bibitem{Lashkarimod}
  N.~Lashkari,
  Phys.\ Rev.\ Lett.\  {\bf 117} (2016) no.4,  041601
  doi:10.1103/PhysRevLett.117.041601
  [arXiv:1508.03506 [hep-th]].

\bibitem{Sarosi}
  G.~Sarosi and T.~Ugajin,
  JHEP {\bf 1702} (2017) 060
  doi:10.1007/JHEP02(2017)060
  [arXiv:1611.02959 [hep-th]];\\
  G.~Sarosi and T.~Ugajin,
  JHEP {\bf 1801} (2018) 012
  doi:10.1007/JHEP01(2018)012
  [arXiv:1705.01486 [hep-th]].

\bibitem{Pando}
  G.~Wong, I.~Klich, L.~A.~Pando Zayas and D.~Vaman,
  JHEP {\bf 1312} (2013) 020
  doi:10.1007/JHEP12(2013)020
  [arXiv:1305.3291 [hep-th]].

\bibitem{Liu}
  N.~Lashkari, H.~Liu and S.~Rajagopal,
  \emph{Modular Flow of Excited States,}
  arXiv:1811.05052 [hep-th].

\bibitem{Peter 1}
  M.~Botta-Cantcheff, P.~Martinez and G.~A.~Silva,
  JHEP {\bf 1602} (2016) 171
  doi:10.1007/JHEP02(2016)171
  [arXiv:1512.07850 [hep-th]].

\bibitem{Skenderis}
  A.~Christodoulou and K.~Skenderis,
  JHEP {\bf 1604} (2016) 096
  doi:10.1007/JHEP04(2016)096
  [arXiv:1602.02039 [hep-th]].
  
\bibitem{Marolf}
  D.~Marolf, O.~Parrikar, C.~Rabideau, A.~Izadi Rad and M.~Van Raamsdonk,
  JHEP {\bf 1806} (2018) 077
  doi:10.1007/JHEP06(2018)077
  [arXiv:1709.10101 [hep-th]].
  
 \bibitem{Peter 2} 
  M.~Botta-Cantcheff, P.~J.~Martinez and G.~A.~Silva,
  JHEP {\bf 1703}, 148 (2017)
  doi:10.1007/JHEP03(2017)148
  [arXiv:1703.02384 [hep-th]].  
  
\bibitem{Belin}
  A.~Belin, A.~Lewkowycz and G.~Sarosi,
  Phys.\ Lett.\ B {\bf 789} (2019) 71
  doi:10.1016/j.physletb.2018.10.071
  [arXiv:1806.10144 [hep-th]]; \\
  A.~Belin, A.~Lewkowycz and G.~Sarosi,
  JHEP {\bf 1903} (2019) 044
  doi:10.1007/JHEP03(2019)044
  [arXiv:1811.03097 [hep-th]].
  
  \bibitem{Mark}
  H.~Z.~Chen and M.~Van Raamsdonk,
  JHEP {\bf 1908} (2019) 062
  doi:10.1007/JHEP08(2019)062
  [arXiv:1903.00972 [hep-th]]; 
  
  \bibitem{Peter 3}
  M.~Botta-Cantcheff, P.~J.~Martinez and G.~A.~Silva,
  JHEP {\bf 1811} (2018) 129
  doi:10.1007/JHEP11(2018)129
  [arXiv:1808.10306 [hep-th]].
  
    \bibitem{Peter 4}
   M.~Botta-Cantcheff, P.~J.~Martinez and G.~A.~Silva,
  JHEP {\bf 1904}, 028 (2019)
  doi:10.1007/JHEP04(2019)028
  [arXiv:1901.00505 [hep-th]].
  
   \bibitem{SvRC}
  K.~Skenderis and B.~C.~van Rees,
  Phys.\ Rev.\ Lett.\  {\bf 101}, 081601 (2008),
  [arXiv:0805.0150 [hep-th]].


\bibitem{SvRL}
  K.~Skenderis and B.~C.~van Rees,
  JHEP {\bf 0905} (2009) 085,
  [arXiv:0812.2909 [hep-th]].
   
   \bibitem{HH}
J.~B.~Hartle and S.~W.~Hawking,
Adv. Ser. Astrophys. Cosmol. \textbf{3}, 174-189 (1987)
doi:10.1103/PhysRevD.28.2960

\bibitem{ensayo-us}
  M.~Botta-Cantcheff and P.~J.~Martinez,
  \emph{Which quantum states are dual to classical spacetimes?,}
  arXiv:1703.03483 [hep-th].

 \bibitem{Vanraam2019} 
  F.~M.~Haehl, E.~Mintun, J.~Pollack, A.~J.~Speranza and M.~Van Raamsdonk,
  JHEP {\bf 1906} (2019) 005
  doi:10.1007/JHEP06(2019)005
  [arXiv:1904.01584 [hep-th]].

\bibitem{Glauber} Roy J. Glauber, Phys. Rev. 84, 395 Published 1 November 1951; \\Roy J. Glauber, Phys. Rev. 130, 2529 Published 15 June 1963


\bibitem{Aitor2013}
  T.~Faulkner, A.~Lewkowycz and J.~Maldacena,
  JHEP {\bf 1311} (2013) 074
  doi:10.1007/JHEP11(2013)074
  [arXiv:1307.2892 [hep-th]].
  
  \bibitem{Faulkner2017}
  T.~Faulkner and A.~Lewkowycz,
  JHEP {\bf 1707} (2017) 151
  doi:10.1007/JHEP07(2017)151
  [arXiv:1704.05464 [hep-th]].

 \bibitem{BDHM}
T. Banks, M. R. Douglas, G. T. Horowitz, and E. J. Martinec,
\emph{AdS dynamics from conformal field theory},
[hep-th/9808016].

\bibitem{eternal}  J.~M.~Maldacena,
  JHEP {\bf 0304}, 021 (2003)
  doi:10.1088/1126-6708/2003/04/021
  [hep-th/0106112].
  
\bibitem{Schwinger}
  J.~S.~Schwinger,
  J.\ Math.\ Phys.\  {\bf 2} (1961) 407.
  doi:10.1063/1.1703727

\bibitem{Keldysh}
  L.~V.~Keldysh,
  Zh.\ Eksp.\ Teor.\ Fiz.\  {\bf 47} (1964) 1515
   [Sov.\ Phys.\ JETP {\bf 20} (1965) 1018].


\bibitem{UmezawaTFD} H.~Umezawa,
  {\it Advanced field theory: Micro, macro, and thermal physics,}
  New York, USA: AIP (1993).
  
  
  \bibitem{UmezawaTFD2}  
Y.~Takahashi and H.~Umezawa, 
  Int.\ J.\ Mod.\ Phys.\ B {\bf 10}, 1755 (1996),
  doi:10.1142/S0217979296000817.

\bibitem{UmezawaInteraction}
  Thermo Field Dynamics in Interaction Representation 
H. Matsumoto, Y. Nakano, H. Umezawa, F. Mancini, M. Marinaro
Progress of Theoretical Physics, Volume 70, Issue 2, August 1983, Pages 599 602,

 \bibitem{BDHM2}
 D.~Harlow and D.~Stanford,
 \emph{Operator Dictionaries and Wave Functions in AdS/CFT and dS/CFT,}
  arXiv:1104.2621 [hep-th].
  
\bibitem{Parrikar2020}
  S.~Balakrishnan and O.~Parrikar,
  \emph{Modular Hamiltonians for Euclidean Path Integral States,}
  arXiv:2002.00018 [hep-th].
  
  \bibitem{Takashi} Y.
Takahashi and H. Umezawa, Collective Phenomena 2 55 (1975)

  \bibitem{Thermal-Coherent} Oz-Vogt, J., Mann, A., and Revzen, M., 
Journal of Modern Optics {\bf38} (1991)  2339-2347.

  \bibitem{Faulkner:2013ica} 
  T.~Faulkner, M.~Guica, T.~Hartman, R.~C.~Myers and M.~Van Raamsdonk,
  JHEP {\bf 1403}, 051 (2014)
  doi:10.1007/JHEP03(2014)051
  [arXiv:1312.7856 [hep-th]].
  
  \bibitem{Reeh}
Reeh, H., Schlieder, S. 
 Nuovo Cim 22, 1051-1068 (1961). https://doi.org/10.1007/BF02787889
  
  \bibitem{Van17}
  T.~Faulkner, F.~M.~Haehl, E.~Hijano, O.~Parrikar, C.~Rabideau and M.~Van Raamsdonk,
  JHEP {\bf 1708}, 057 (2017)
  doi:10.1007/JHEP08(2017)057
  [arXiv:1705.03026 [hep-th]].
  
  \bibitem{BootsBrazil2} M.~Botta Cantcheff, A.~L.~Gadelha, D.~F.~Z.~Marchioro and D.~L.~Nedel,
  Eur.\ Phys.\ J.\ C {\bf 78}, no. 2, 105 (2018)
  doi:10.1140/epjc/s10052-018-5545-2
  [arXiv:1702.02069 [hep-th]].

  \bibitem{Kenmoku}
  M. Kenmoku, M. Kuwata and K. Shigemoto, Class. Quant. Grav. 25, 145016 (2008) doi:10.1088/0264-
9381/25/14/145016 [arXiv:0801.2044 [gr-qc]].

\bibitem{Boots-StringTFD}	
 M.~Botta Cantcheff,
  Eur.\ Phys.\ J.\ C {\bf 55}, 517 (2008)
  doi:10.1140/epjc/s10052-008-0603-9
  [arXiv:0710.3186 [hep-th]].
  
  
\bibitem{Dong2018}
  X.~Dong, D.~Harlow and D.~Marolf,
  JHEP {\bf 1910} (2019) 240
  doi:10.1007/JHEP10(2019)240
  [arXiv:1811.05382 [hep-th]].

\bibitem{botta}
  M.~Botta Cantcheff,
  \emph{Area Operators in Holographic Quantum Gravity},
  arXiv:1404.3105 [hep-th].

\bibitem{Hayward19}
  T.~Takayanagi and K.~Tamaoka,
 \emph{Gravity Edges Modes and Hayward Term,}
  arXiv:1912.01636 [hep-th].
  
\bibitem{Hayward1993}
  G.~Hayward,
  Phys.\ Rev.\ D {\bf 47} (1993) 3275.
  doi:10.1103/PhysRevD.47.3275
  
  \bibitem{Botta-Hayward}
M.~Botta-Cantcheff, P.~J.~Martinez and J.~F.~Zarate,
[arXiv:2005.11338 [hep-th]], to appear in JHEP.
  
  
 \bibitem{Israel}
W. Israel, Phys. Lett. A 57, 107 (1976).


\bibitem{RindlerAdSCFT}
M.~Parikh and P.~Samantray,
JHEP \textbf{10}, 129 (2018)
doi:10.1007/JHEP10(2018)129
[arXiv:1211.7370 [hep-th]].

\bibitem{TFDyTT}
 N.~P.~Landsman and C.~G.~van Weert,
  Phys.\ Rept.\  {\bf 145} (1987) 141.
  doi:10.1016/0370-1573(87)90121-9
 \end{thebibliography}
 \end{document}